\documentclass[aps,prd,twocolumn,showpacs,superscriptaddress]{revtex4}
\usepackage{graphics}
\usepackage{amssymb}
\usepackage{amsmath}
\usepackage{epsfig}
\usepackage{color}
\usepackage{verbatim}
\usepackage{float}
\usepackage{mathrsfs}
\usepackage{esint}
\usepackage{hyperref}

\begin{document}
\title{Fulde-Ferrell states and Berezinskii-Kosterlitz-Thouless phase transition in two-dimensional imbalanced Fermi gases}

\author{Shaoyu Yin}
\address{COMP Centre of Excellence, Department of Applied Physics, Aalto University, FI-00076 Aalto, Finland}
\author{J.-P. Martikainen}
\address{COMP Centre of Excellence, Department of Applied Physics, Aalto University, FI-00076 Aalto, Finland}
\author{P. T\"orm\"a\footnote{paivi.torma@aalto.fi}}
\address{COMP Centre of Excellence, Department of Applied Physics, Aalto University, FI-00076 Aalto, Finland}
\address{Kavli Institute for Theoretical Physics, University of California, Santa Barbara, California 93106-4030, USA}

\begin{abstract}
We study the superfluid properties of two-dimensional spin-population-imbalanced Fermi gases to explore the interplay between the Berezinskii-Kosterlitz-Thouless~(BKT) phase transition and the possible instability towards the Fulde-Ferrell~(FF) state. By the mean-field approximation together with quantum fluctuations, we obtain phase diagrams as functions of temperature, chemical potential imbalance and binding energy. We find that the fluctuations change the mean-field phase diagram significantly. We also address possible effects of the phase separation and/or the anisotropic FF phase to the BKT mechanism. The superfluid density tensor of the FF state is obtained, and its transverse component is found always vanishing. This causes divergent fluctuations and possibly precludes the existence of the FF state at any non-zero temperature.
\end{abstract}
\maketitle

\section{introduction}

Systems at low temperature can exhibit diverse quantum mechanical phenomena, such as superfluidity, superconductivity, Bose-Einstein condensation~(BEC), Mott insulators, and various magnetic states. Such phenomena become possible because of the interplay between interactions and low temperature. Fast progress in the ultracold-gas experiments (see e.g.~\cite{Bloch:2008} and references therein) has made these highly controllable systems attractive for the study of correlated quantum states. While remarkable experimental achievements have been reached on different quantum states in various settings, one state of special interest, namely the inhomogeneous superfluidity with non-constant order parameter remains as a challenge. Such possibility has been predicted on spin-population-imbalanced Fermi systems several decades ago~\cite{FF,LO}. In such a state Cooper pairs can have non-zero total momenta. 

The simplest case of inhomogeneous superfluidity is the Fulde-Ferrell~(FF) state~\cite{FF}, where the order parameter is a single plane wave, $\Delta_0e^{i2\mathbf{Q}\cdot\mathbf{x}}$, with $\Delta_0$ being the magnitude of the order parameter and $2\mathbf{Q}$ as the momentum of the pair (sometimes referred as the FF(LO) vector). One may also consider the Larkin-Ovchinnikov~(LO) state~\cite{LO} with $\Delta_0\cos(2\mathbf{Q}\cdot\mathbf{x})$, which can be taken as the superposition of two equal FF modes with opposite momenta \cite{Radzihovsky:2011}. More generally, the nonuniform order parameter can be expressed as a superposition of many possible FFLO vectors by $\sum_\mathbf{Q}\Delta_{0\mathbf{Q}}e^{i2\mathbf{Q}\cdot\mathbf{x}}$. All these states are usually categorized as FFLO states and have been extensively studied. Although an undisputed experimental evidence is still missing, there have already been several experiments in heavy fermion superconductors~\cite{Radovan:2003,Bianchi:2003,Won:2004,Watanabe:2004,Capan:2004,Martin:2005,Kakuyanagi:2005,Kumagai:2006,Correa:2007} and organic superconductors~\cite{Lortz:2007,Coniglio:2010} which report signatures consistent with the predicted FFLO states. The recent realization of imbalanced Fermi gases with ultra-cold atoms~\cite{Zwierlein:2006,Partridge:2006} has triggered more interest in the FFLO states. An experiment with a one-dimensional~(1D) ultracold Fermi gas showed results consistent with the FFLO state~\cite{Liao:2010}, but direct observation, especially in higher dimensions, remains as a goal.

Besides physical parameters such as temperature $T$, particle density, and interaction strength, dimensionality may also affect the properties of the quantum systems significantly. It is well known that thermal fluctuations become increasingly strong as dimensionality is lowered. The Mermin-Wagner-Hohenberg theorem states clearly that there cannot be any long-range order in uniform 1D or two-dimensional~(2D) systems at $T\neq0$~\cite{MWH}. However, the 2D case turns out to be marginal and a quasi-long-range order can survive at low temperatures in the presence of interactions or trapping potential. This suggests that 2D systems can display very rich phenomena~\cite{Esslinger:2006}.

One peculiar possibility in 2D systems is the Berezinskii-Kosterlitz-Thouless~(BKT) phase transition~\cite{Berezinskii:1971,KT}. It describes a mechanism by which the quasi-long-range order of a 2D system is destroyed by the proliferation of free vortices and antivortices when the temperature is higher than a critical value $T_\mathrm{BKT}$. Below $T_\mathrm{BKT}$ the quasi-long-range order is sufficient for the existence of superfluidity. Furthermore, it has been theoretically shown that a 2D quantum gas can also form a BEC in the presence of a trapping potential~\cite{Bagnato:1991}. 

Since the properties of 2D Fermi gases can be related to other important (quasi-)~2D systems, such as graphene~\cite{Beenakker:2008}, and the 2D CuO$_2$ layers which play a significant role in high $T_c$ superconductors~\cite{Dagotto:1994}, their scientific importance extends beyond the field of ultra-cold gases. There have been some theoretical studies of various properties of the 2D imbalanced Fermi gases~\cite{Tempere:2007,He:2008,Tempere:2009,Klimin:2011,Du:2012,Klimin:2012}. Here we study the possibility and properties of the FFLO phase in a 2D imbalanced Fermi gas, especially the interplay between FFLO states, phase separation, and the BKT phase transition. A similar question was posed briefly in a letter by H. Shimahara~\cite{Shimahara:1998} in the context of a 2D superconductor based on the Ginzburg-Landau~(GL) theory, but the anisotropic superfluid density (stiffness) was not taken into account. Recently for imbalanced Fermi gases, the GL theory has been applied to the study of the LO state in various dimensions \cite{Radzihovsky:2009}.

In the present paper we discuss this topic by using mean-field~(MF) theory with fluctuations. Our discussion is not limited to a small order parameter and goes beyond the GL theory. Fair amount of relevant theoretical work devoted to the research of FFLO states has been published under different conditions and various dimensions. For three-dimensional~(3D) homogeneous imbalanced Fermi gases, the FF state is expected to exist in a narrow sliver in the phase diagram~\cite{Sheehy:dual,Parish:2007}. In isotropic traps, FFLO features are predicted to appear only as a boundary layer \cite{Kinnunen:2006,Jensen:2007}, although highly anisotropic traps yield much larger FFLO phase areas~\cite{Machida:2006,Kim:2011}. Interestingly, in optical lattices the FFLO state has been suggested to be stabilized due to nesting of the Fermi surfaces~\cite{Koponen:2007PRL,Koponen:2007NJP,Loh:2010}. For the case of (quasi-)~1D system, where no long-range order exists due to extremely strong fluctuations, the possibility of FFLO state was first discussed in the context of superconductors by using the bosonization of electron gases \cite{Yang:2001}, and later for atomic gases many numerical simulations show the existence of the FFLO state~\cite{Feiguin:2007,Tezuka:2008,Batrouni:2008,Rizzi:2008}, which is also supported by a few solvable models~\cite{Machida:1984,Machida:2005,Orso:2007,Hu:2007,Zhao:2008}, and several methods have been proposed for the detection of such 1D FFLO states~\cite{Bakhtiari:2008,Korolyuk:2010,Kajala:2011,Chen:2012,Lu:2012}. However, the (quasi-)~2D imbalanced case with quasi-long-range order is less explored because of its complexity, especially the marginally strong fluctuations. Some lattice simulations show that the FF state exists with medium filling factor, but it is unclear what happens in the zero-filling-factor limit, i.e. the continuum limit~\cite{Koponen:2007NJP}. Because of the recent progress in ultra-cold atoms, especially the realization of degenerate quasi-2D atomic gases both for bosons~\cite{Hadzibabic:2006} and fermions~\cite{Martiyanov:2010} by using 1D optical lattices with lattice depths $V_0$ in the range of $V_0/h\approx10\cdots100$~kHz (here $h$ is the Planck constant), many important properties of 2D systems have been observed. For Fermi systems, these include studies of pseudogap physics~\cite{Feld:2011} and polarons in imbalanced gases~\cite{Koschorreck:2012}. These ground-breaking experiments provide a strong motivation to address the issue of polarized 2D Fermi gases with the possibility of the FF state. Although recent experiments usually study the quasi-2D gases, for the sake of simplicity, we will focus only on the perfect 2D case which corresponds to the limit of an infinitely deep trapping in the third dimension. Therefore, we will not discuss some interesting phenomena such as the FFLO states in a dimensional crossover \cite{Kim:2012,Sun:2013,Heikkinen:2013}. It is also worth mentioning that, in the opposite limit, i.e. with a very loose trap in the third direction, a 3D gas with 1D periodic potential not only stabilizes the possible FFLO states but also enables the FFLO wavevector to lie skewed with respect to the potential \cite{Devreese:2011}.

\begin{figure}
\includegraphics[width=0.78\columnwidth]{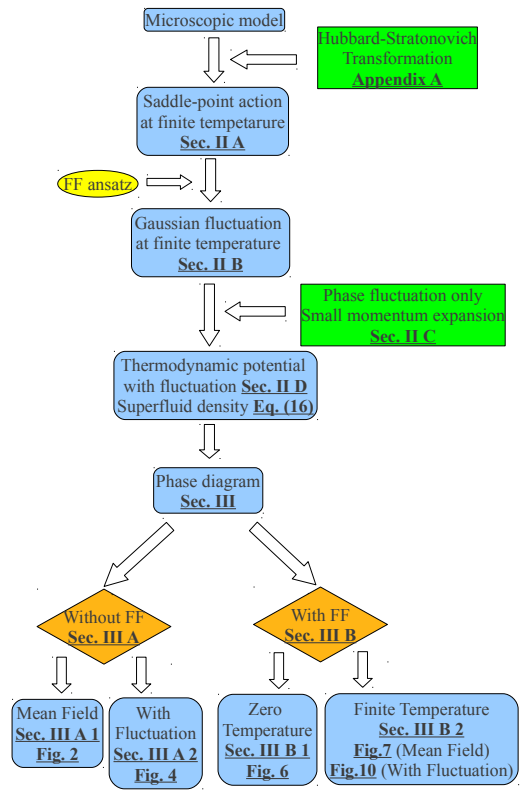}
\caption{(Color online) Framework of the paper, where the yellow oblate indicates a use of an ansatz, the green rectangles indicate approximations, 
while orange diamonds imply different applications of the theory. The related sections and important equations and figures are indicated by underlined boldface font.}\label{flowchart}
\end{figure}

This paper is organized as follows. We start, in Sec.~\ref{Sec-action}, with a MF approximation by calculating the saddle-point action of the system. Since fluctuations are not negligible in a 2D system, the fluctuation contribution is included in Sec.~\ref{Sec-fluctuation}. Based on these results we can proceed, in Sec.~\ref{Sec-Omega}, to minimize the total thermodynamic potential and examine the phase diagram and possible phase transitions in Sec.~\ref{Sec-phasediagram}. For the sake of simplicity, in the present paper we focus on the FF state. Since it is commonly accepted that the LO state is usually more stable and energetically favorable than the FF state, stability of FF indicates stability of LO as well. We summarize the structure of the paper in the flowchart of Fig.~\ref{flowchart}. Throughout this paper we use the natural units with $\hbar=k_B=1$. Some notations are defined in the beginning of Appendix~\ref{transformation}.

\section{Theoretical Framework}

\subsection{Saddle-Point Action}\label{Sec-action}

We assume a system of fermions with two species, namely, spin up ($\sigma=\uparrow$) and spin down ($\sigma=\downarrow$). Hamiltonian density in terms of the creation $\hat\psi^\dagger_\sigma(x)$ and annihilation operators $\hat\psi_\sigma(x)$ reads
\[\hat H(x)=\sum_\sigma\hat\psi^\dagger_\sigma(x)(\hat\varepsilon-\mu_\sigma)\hat\psi_\sigma(x)-g\hat\psi^\dagger_\uparrow(x)\hat\psi^\dagger_\downarrow(x)\hat\psi_\downarrow(x)\hat\psi_\uparrow(x).\]
Here $\hat\varepsilon$ is the kinetic energy operator and $\mu_\sigma$ the chemical potential for spin $\sigma$ (from which we define $\mu=(\mu_\uparrow+\mu_\downarrow)/2$ and $h=(\mu_\uparrow-\mu_\downarrow)/2$ for later convenience), $g>0$ is the strength of the attractive contact interaction. By using the standard Hubbard-Stratonovich transformation (cf. Appendix~\ref{transformation}) with the auxiliary field operator $\hat\Delta$ coupled to $\hat\psi^\dagger_\uparrow\hat\psi^\dagger_\downarrow$, we can obtain the effective action
\begin{align}\label{effaction}
S_\mathrm{eff}=\mathcal{V}\sum_{iq_n,\mathbf{q}}\frac{|\hat\Delta(q)|^2}{g}-\mathrm{Tr}\ln[\beta\mathbf{G}^{-1}(k,k')],
\end{align}
where $\mathcal{V}=V\beta$ with $V$ as the volume and $\beta$ as the inverse of temperature $T$, $k$ (as well as $q$) includes both the Matsubara frequency $ik_n$ and the vector space momentum $\mathbf{k}$, and the inverse of the Nambu propagator $\mathbf{G}^{-1}(k,k')$ is a $2\times2$ matrix in the Nambu space given by
\[\left(\begin{array}{cc}
(ik'_n-\epsilon_\mathbf{k'}+\mu_\uparrow)\delta_{k,k'} & \hat\Delta(k-k')\\
\hat\Delta^*(-k+k') & (ik'_n+\epsilon_\mathbf{k'}-\mu_\downarrow)\delta_{k,k'}
\end{array}\right).\]
Here $\epsilon_\mathbf{k}$ is the kinetic energy of a particle with momentum $\mathbf{k}$, and $\mathrm{Tr}$ means the trace over the Nambu space, the momentum space, and the Matsubara frequencies.

In the MF approximation, the field operator $\hat\Delta$ is replaced by its saddle-point value, namely the order parameter $\Delta_\mathrm{s}$, which satisfies $\partial S_\mathrm{eff}/\partial\Delta^*_\mathrm{s}=0$. In the case of balanced Fermi gases, the momenta of the paired fermions are equal in magnitude but with opposite directions, such that $\Delta_\mathrm{s}$ is a constant. However with imbalance, the pairs might have non-zero momenta, which results in the FF(LO) states.

Here we examine the FF state with $\Delta_\mathrm{s}=\Delta_0e^{2i\mathbf{Q}\cdot\mathbf{x}}$. Its phase part can be absorbed by a momentum shift in the corresponding Fermi fields $\hat\psi_\sigma$ (cf. Appendix~\ref{transformation}), yielding $\tilde\Delta_\mathrm{s}=\Delta_0$ and a new Nambu propagator $\tilde{\mathbf{G}}_\mathrm{s}^{-1}(k,k')=\tilde{\mathbf{G}}_\mathrm{s}^{-1}(k)\delta_{k,k'}$ which is diagonal in momentum space, and
\begin{equation}
\tilde{\mathbf{G}}_\mathrm{s}^{-1}(k)=\left(\begin{array}{cc}
ik_n-\epsilon_\mathbf{Q+k}+\mu_\uparrow & \Delta_0\\
\Delta_0 & ik_n+\epsilon_\mathbf{Q-k}-\mu_\downarrow
\end{array}\right),\label{shiftpropinv}
\end{equation}
which can be straightforwardly inverted as
\begin{align}
\tilde{\mathbf{G}}_\mathrm{s}(k)=&\frac{1}{(ik_n-\epsilon_\mathbf{Q+k}+\mu_\uparrow)(ik_n+\epsilon_\mathbf{Q-k}-\mu_\downarrow)-\Delta_0^2}\nonumber\\ &\times\left(\begin{array}{cc}
ik_n+\epsilon_\mathbf{Q-k}-\mu_\downarrow & -\Delta_0\\
-\Delta_0 & ik_n-\epsilon_\mathbf{Q+k}+\mu_\uparrow\end{array}\right).
\end{align}
It is useful to note that the denominator of $\tilde{\mathbf{G}}_\mathrm{s}$ is simply $\mathrm{det}(\tilde{\mathbf{G}}_\mathrm{s}^{-1})=1/\mathrm{det}(\tilde{\mathbf{G}}_\mathrm{s})$. Substituting $\tilde\Delta_\mathrm{s}=\Delta_0$ and $\tilde{\mathbf{G}}_\mathrm{s}^{-1}$ into Eq.~(\ref{saddle}), we get the saddle-point action which reads, after Matsubara summation,
\begin{equation}\label{saddleaction}
S_\mathrm{s}=\frac{\mathcal{V}\Delta_0^2}{g}-\sum_\mathbf{k}\{\ln[2\cosh(\beta E_\mathbf{Qk})+2\cosh(\beta h_\mathbf{Qk})]-\beta\xi_\mathbf{Qk}\},
\end{equation}
where $\xi_\mathbf{Qk}=\frac{\mathbf{Q}^2+\mathbf{k}^2}{2m}-\mu$, $E_\mathbf{Qk}=\sqrt{\xi_\mathbf{Qk}^2+\Delta_0^2}$, and $h_\mathbf{Qk}=h-\frac{\mathbf{Q}\cdot\mathbf{k}}{m}$. Here a quadratic dispersion is assumed for concreteness.

\subsection{Fluctuations}\label{Sec-fluctuation}

In order to go beyond the MF approximation, we introduce fluctuations to the order parameter. Conventionally, for the study of the 2D BKT phase transition, it is convenient to work with a phase fluctuation via $\Delta\rightarrow\Delta_0e^{i\theta(x)}$. More generally we could have $\Delta\rightarrow(\Delta_0+\eta(x))e^{i\theta(x)}$, where two real fields $\eta(x)$ and $\theta(x)$ represent the amplitude and the phase fluctuations, respectively. For the FF ansatz, we use $(\Delta_0+\eta(x))e^{2i\mathbf{Q}\cdot\mathbf{x}+i\theta(x)}$, such that $\theta(x)$ fluctuates around the phase of the FF saddle-point ansatz.

Notice that while $\theta(x)$ is not necessarily small, its derivatives can be taken as small perturbative parameters since we can expect a smooth phase change of the order parameter in the space-time when $T$ is not very high and the fluctuation picture is valid. For this reason, it is more convenient to start the derivation in the coordinate space rather than in the momentum space. Also, in order to separate the perturbative part in $\mathbf{G}^{-1}$ more easily, we first apply a phase rotation to the Nambu basis to absorb the phase of $\Delta$~\cite{Diener:2008} by the transformation
\begin{equation}\label{gaugetransform}
\hat\Psi(x)\rightarrow\tilde{\hat\Psi}(x)=U(x)\hat\Psi(x),
\end{equation}
with
\[U(x)=\left(\begin{array}{cc}
e^{-i\mathbf{Q}\cdot\mathbf{x}-i\theta(x)/2} & 0\\
0 & e^{i\mathbf{Q}\cdot\mathbf{x}+i\theta(x)/2}
\end{array}\right).\]
This is a generalization of the momentum shift we used in Sec.~\ref{Sec-action} to get $\tilde{\mathbf{G}}_\mathrm{s}^{-1}$. Note that there is no mixing between the two fields of different species since $U$ is diagonal. Correspondingly
\begin{widetext}
\begin{align}
\tilde{\mathbf{G}}^{-1}(x,x')&=U(x)\mathbf{G}^{-1}(x,x')U^\dagger(x')=\left(\begin{array}{cc}
-\frac{i}{2}\partial_\tau\theta-\partial_\tau-\hat\varepsilon_{\mathbf{Q}+\frac{\nabla\theta}{2}}+\mu_\uparrow & \Delta_0+\eta(x)\\
\Delta_0+\eta(x) & \frac{i}{2}\partial_\tau\theta-\partial_\tau+\hat\varepsilon_{-\mathbf{Q}-\frac{\nabla\theta}{2}}-\mu_\downarrow
\end{array}\right)\delta(x-x'),
\end{align}
\end{widetext}
where $\hat\varepsilon_{\pm(\mathbf{Q}+\frac{\nabla\theta}{2})}$ means the momentum of the energy operator is shifted by $\pm(\mathbf{Q}+\frac{\nabla\theta}{2})$, e.g. $\hat\varepsilon_{\pm(\mathbf{Q}+\frac{\nabla\theta}{2})}f(\mathbf{k})=f(\mathbf{k})\epsilon_{\mathbf{k}\pm(\mathbf{Q}+\frac{\nabla\theta}{2})}$. Meanwhile, the order parameter becomes $\tilde\Delta(x)=\Delta_0+\eta(x)$ with the Fourier transform 
\begin{equation}\label{deltaq}
\tilde\Delta(q)=\Delta_0\delta_{q,0}+\eta(q).
\end{equation}

Now we can separate out a perturbative matrix $\tilde{\mathbf{K}}$ from $\tilde{\mathbf{G}}^{-1}=\tilde{\mathbf{G}}^{-1}_\mathrm{s}+\tilde{\mathbf{K}}$ with $\eta$ and $\nabla\theta$ as small variables, where \[\tilde{\mathbf{G}}^{-1}_\mathrm{s}=\left(\begin{array}{cc}
-\partial_\tau-\hat\varepsilon_\mathbf{Q}+\mu_\uparrow & \Delta_0\\
\Delta_0 & -\partial_\tau+\hat\varepsilon_{-\mathbf{Q}}-\mu_\downarrow
\end{array}\right)\delta(x-x')\]
is the Fourier transform of Eq.~(\ref{shiftpropinv}), while
\begin{widetext}
\begin{align}
\tilde{\mathbf{K}}(x,x')&=\left(\begin{array}{cc}
-\frac{i}{2}\partial_\tau\theta-\hat\varepsilon_{\mathbf{Q}+\frac{\nabla\theta}{2}}+\hat\varepsilon_\mathbf{Q} & \eta(x)\\
\eta(x) & \frac{i}{2}\partial_\tau\theta+\hat\varepsilon_{-\mathbf{Q}-\frac{\nabla\theta}{2}}-\hat\varepsilon_{-\mathbf{Q}}
\end{array}\right)\delta(x-x')\\
&=\left(\begin{array}{cc}
-\frac{i}{2}\partial_\tau\theta+\frac{i}{2m}(\nabla\theta\cdot\nabla_\mathbf{Q}+\frac{1}{2}\nabla_\mathbf{Q}\cdot\nabla\theta)-\frac{(\nabla\theta)^2}{8m} & \eta(x)\\
\eta(x) & \frac{i}{2}\partial_\tau\theta+\frac{i}{2m}(\nabla\theta\cdot\nabla_\mathbf{-Q}+\frac{1}{2}\nabla_\mathbf{-Q}\cdot\nabla\theta)+\frac{(\nabla\theta)^2}{8m}
\end{array}\right)\delta(x-x').\nonumber
\end{align}
\end{widetext}
Here in the last line we separated the perturbative $\nabla\theta$ from the non-relativistic dispersion $\hat\varepsilon_{\pm(\mathbf{Q}+\frac{\nabla\theta}{2})}\equiv-\frac{\nabla_{\pm(\mathbf{Q}+\frac{\nabla\theta}{2})}^2}{2m}$. Note that our derivation was quite general until this point and most of it is equally valid, for example, in optical lattices with a different dispersion. From here on our formulae apply only in homogeneous space because of the specific quadratic dispersions.

The Fourier transform of $\tilde{\mathbf{K}}(x,x')$ is
\begin{widetext}
\begin{equation}\label{perturbationK}
\tilde{\mathbf{K}}(k,k')=\sum_q\left[\eta(q)\sigma_1-\frac{q_n\theta(q)}{2}\sigma_3-\frac{i\theta(q)}{4m}(\mathbf{k}^2-\mathbf{k'}^2+3\mathbf{q}\cdot\mathbf{Q}\sigma_3)\right]\delta_{k-k',q}+\sum_{q,q'}\frac{\theta(q)\theta(q')\mathbf{q}\cdot\mathbf{q}'}{8m}\sigma_3\delta_{k-k',q+q'}\equiv\tilde{\mathbf{K}}_1+\tilde{\mathbf{K}}_2,
\end{equation}
\end{widetext}
where the Pauli matrices $\sigma_1=\left(\begin{matrix}0&\ 1\\1&\ 0\end{matrix}\right)$ and $\sigma_3=\left(\begin{matrix}1&\ 0\\0&\ -1\end{matrix}\right)$ operating in the Nambu space were introduced to make expressions more compact. Besides, as $\mathbf{q}\theta(q)$ corresponds to $\nabla\theta$, in the Fourier transformation sense, and $q_n\theta(q)$ to $\partial_\tau\theta$, we take them as the small parameters of the same order as $\eta(q)$. Therefore in Eq.~(\ref{perturbationK}) the double-sum term labelled as $\tilde{\mathbf{K}}_2$ corresponds to the second-order perturbation, while the remaining part $\tilde{\mathbf{K}}_1$ is the first order perturbation.

Now we can obtain the effective action by using $\tilde{\mathbf{G}}^{-1}(k,k')=\tilde{\mathbf{G}}^{-1}_\mathrm{s}(k)\delta_{k,k'}+\tilde{\mathbf{K}}(k,k')$, with $\tilde{\mathbf{G}}^{-1}_\mathrm{s}(k)$ from Eq.~(\ref{shiftpropinv}) and $\tilde{\mathbf{K}}(k,k')$ from Eq.~(\ref{perturbationK}), inserted into Eq.~(\ref{effaction}) together with $\tilde\Delta(q)$ from Eq.~(\ref{deltaq}). Subtracting the saddle-point action $S_\mathrm{s}=S_\mathrm{eff}(\tilde\Delta_\mathrm{s})=S_\mathrm{eff}(\Delta_0\delta_{q,0})$, we find the fluctuation action
\begin{align}\label{sfl}
S_\mathrm{fl}&=S_\mathrm{eff}(\Delta)-S_\mathrm{s}\\ &=\mathcal{V}\sum_q\frac{\Delta_0\delta_{q,0}\eta^*(q)+\Delta_0\delta_{q,0}\eta(q)+|\eta(q)|^2}{g}\nonumber\\ &\qquad\quad-\mathrm{Tr}\ln[1+\tilde{\mathbf{G}}_\mathrm{s}\tilde{\mathbf{K}}]\nonumber\\ &=\frac{2\mathcal{V}\Delta_0\eta(0)}{g}+\frac{\mathcal{V}\sum_q|\eta(q)|^2}{g}-\sum_{k}\mathrm{tr}\tilde{\mathbf{G}}_\mathrm{s}(k)\tilde{\mathbf{K}}(k,k)\nonumber\\ &\quad\quad+\frac{1}{2}\sum_{k,k'}\mathrm{tr}\tilde{\mathbf{G}}_\mathrm{s}(k)\tilde{\mathbf{K}}_1(k,k')\tilde{\mathbf{G}}_\mathrm{s}(k')\tilde{\mathbf{K}}_1(k',k)+\cdots,\nonumber
\end{align}
where only terms up to the second order are kept. Note that $\tilde{\mathbf{K}}(k,k)=\eta(0)\sigma_1-\sum_q\frac{\theta(q)\theta(-q)\mathbf{q}^2}{8m}\sigma_3=\eta(0)\sigma_1-\sum_q\frac{|\theta(q)|^2\mathbf{q}^2}{8m}\sigma_3$, where the term linear in the perturbative fields is simply $\eta(0)\sigma_1$. With two perturbative fields $\eta$ and $\theta$, the saddle-point condition $(\partial S/\partial\Delta)_{\Delta=\Delta_\mathrm{s}}=0$ requires $\left(\frac{\partial S}{\partial\eta}\right)_{\theta=0}=0$ and $\left(\frac{\partial S}{\partial\theta}\right)_{\eta=0}=0$, where the total action $S=S_\mathrm{s}+S_\mathrm{fl}$. These ensure the vanishing of terms linear in $\eta$ and $\theta$ in the expansion of $S$. Since the linear term of $S_\mathrm{fl}$ is independent of $\theta$, one can obtain only one equation from $\eta$, i.e. constraint on the amplitude of the order parameter. By collecting the terms linear in $\eta$ from Eq.~(\ref{sfl}), we get
\begin{align}
&\frac{2\mathcal{V}\Delta_0\eta(0)}{g}-\sum_k\mathrm{tr}\tilde{\mathbf{G}}_\mathrm{s}(k)\eta(0)\sigma_1\nonumber\\ &\qquad\qquad\qquad\qquad=2\eta(0)\Delta_0\left[\frac{\mathcal{V}}{g}+\sum_k\mathrm{det}\tilde{\mathbf{G}}_\mathrm{s}(k)\right],\nonumber
\end{align}
so the saddle-point condition becomes
\begin{equation}\label{gapequation}
\frac{\mathcal{V}}{g}+\sum_k\mathrm{det}\tilde{\mathbf{G}}_\mathrm{s}(k)=0.
\end{equation}
This result is equivalent to the gap equation which we get by taking the partial derivative of the MF action $S_\mathrm{s}$ with respect to $\Delta_0$. On the other hand, the absence of $\theta$ in the linear expansion of the action means that the saddle-point condition is not enough to determine the phase of the order parameter. We attribute this to the special form of the FF ansatz. As both the FF vector and the phase fluctuation appear in the phase of the order parameter, $i[2\mathbf{Q\cdot x}+\theta(x)]$, it is always possible to redefine $Q$ by separating an arbitrary term linear in $\mathbf{x}$ from $\theta(x)$. This will cause some ambiguity when we determine $Q$, which is to be discussed in detail in Sec.~\ref{Sec-Omega}.

After removing the linear terms according to Eq.~(\ref{gapequation}), we can rewrite Eq.~(\ref{sfl}) in the Gaussian form,
\begin{equation}\label{Gauss-fl-action}
S_\mathrm{fl}=\frac{1}{2}\sum_q(\eta^*(q),\theta^*(q))\mathbf{D}\left(\begin{array}{c}\eta(q)\\\theta(q)\end{array}\right),
\end{equation}
where
\begin{align}\label{Dij}
\mathbf{D}_{11}&=\frac{2\mathcal{V}}{g}+\sum_k\mathrm{tr}\tilde{\mathbf{G}}_\mathrm{s}(k)\sigma_1\tilde{\mathbf{G}}_\mathrm{s}(k+q)\sigma_1,\nonumber\\
\mathbf{D}_{12}&=-\mathbf{D}_{21}=i\sum_k\mathrm{tr}\tilde{\mathbf{G}}_\mathrm{s}(k)\mathbf{J}\tilde{\mathbf{G}}_\mathrm{s}(k+q)\sigma_1,\nonumber\\
\mathbf{D}_{22}&=\sum_k\left[\frac{\mathbf{q}^2}{4m}\mathrm{tr}\tilde{\mathbf{G}}_\mathrm{s}(k)\sigma_3+\mathrm{tr}\tilde{\mathbf{G}}_\mathrm{s}(k)\mathbf{J}\tilde{\mathbf{G}}_\mathrm{s}(k+q)\mathbf{J}\right],
\end{align}
and
\[\mathbf{J}\equiv\frac{iq_n\sigma_3}{2}-\frac{\mathbf{(k+q)}^2-\mathbf{k}^2+3\mathbf{q}\cdot\mathbf{Q}\sigma_3}{4m}.\]
Eqs.~(\ref{Gauss-fl-action}) and (\ref{Dij}) are generalizations of the results of Eq.~(54) in Ref.~\cite{Diener:2008} (we believe the results there were accidentally divided by two twice) to include the possibility of the FF state.

\subsection{Phase Fluctuation and Superfluid Density}\label{Sec-phasefl}

To study the BKT phase transition, it is customary to include only the phase fluctuation and therefore set $\eta=0$. As a result, we now focus only on $\mathbf{D}_{22}$ (cf. the form of Eq.~(\ref{Gauss-fl-action})). Its Matsubara summation is complicated, however, when the phase fluctuation is smooth enough the momentum $q$ can be taken as a small parameter. Since $\mathbf{D}_{22}$ vanishes at the low-frequency and long-wavelength limit, i.e. $iq_n\rightarrow0$ and $\mathbf{q}\rightarrow0$, we expand the fluctuation action Eq.~(\ref{Gauss-fl-action}) with only $\mathbf{D}_{22}\neq0$ and keep the leading (quadratic) order of $q$, and get an approximation for $S_\mathrm{fl}$ as
\begin{equation}\label{flaction}
S_\mathrm{w}=\frac{\mathcal{V}}{2}\sum_q(\kappa q_n^2+\tilde\rho_{ij}q_iq_j)|\theta(q)|^2.
\end{equation}
The expressions for $\kappa$ and $\tilde\rho_{ij}$ are (for an equivalent derivation based on the direct expansion of the saddle-point action, cf. Appendix.~\ref{action-fl})
\begin{align}
\kappa=&\frac{1}{V}\sum_\mathbf{k}\frac{\Delta_0^2X_\mathbf{k}+\beta E_\mathbf{Qk}\xi_\mathbf{Qk}^2Y_\mathbf{k}}{4E_\mathbf{Qk}^3},\label{kappa}\\
\tilde\rho_{ij}=&\frac{1}{V}\sum_\mathbf{k}\left[\frac{\delta_{ij}}{4m}\left(1-\frac{\xi_\mathbf{Qk}}{E_\mathbf{Qk}}X_\mathbf{k}\right)-\frac{\beta Y_\mathbf{k}k_i^2\delta_{ij}}{4m^2}\right.\nonumber\\&\left.\qquad\qquad-3Z_\mathbf{k}k_zQ\delta_{iz}\delta_{jz}\vphantom{\frac{1}{2}}\right]-\frac{9\kappa Q^2\delta_{iz}\delta_{jz}}{4m^2},\label{rho}
\end{align}
where the direction of $\mathbf{Q}$ is chosen as the z-axis, and
\begin{align}
X_\mathbf{k}&\equiv\frac{\sinh(\beta E_\mathbf{Qk})}{\cosh(\beta E_\mathbf{Qk})+\cosh(\beta h_\mathbf{Qk})},\nonumber\\
Y_\mathbf{k}&\equiv\frac{1+\cosh(\beta E_\mathbf{Qk})\cosh(\beta h_\mathbf{Qk})}{[\cosh(\beta E_\mathbf{Qk})+\cosh(\beta h_\mathbf{Qk})]^2},\nonumber
\end{align}
and
\[Z_\mathbf{k}\equiv\frac{\beta\xi_\mathbf{Qk}}{4E_\mathbf{Qk}m^2}\frac{\sinh(\beta E_\mathbf{Qk})\sinh(\beta h_\mathbf{Qk})}{[\cosh(\beta E_\mathbf{Qk})+\cosh(\beta h_\mathbf{Qk})]^2}.\]
Definitions of $\xi_\mathbf{Qk}$, $E_\mathbf{Qk}$ and $h_\mathbf{Qk}$ were given after Eq.~(\ref{saddleaction}).

$S_\mathrm{w}$ describes a Bose gas of spin waves with an anisotropic superfluid density tensor $\tilde\rho_{ij}$. As we know from the isotropic case where $\tilde\rho_{ij}=\rho_0\delta_{ij}$, the spin-wave contribution to the action has a spectrum $\omega_\mathrm{w}(\mathbf{q})=v_\mathrm{w}|\mathbf{q}|$ with the wave speed $v_\mathrm{w}=\sqrt{\rho_0/\kappa}$, and the thermodynamic potential $\Omega_\mathrm{w}=\frac{1}{\mathcal{V}}\sum_\mathbf{q}\ln[1-e^{-\beta\omega_\mathrm{w}(\mathbf{q})}]$~\cite{Botelho:2006}. The BKT transition temperature is determined by~\cite{KT,Nelson:1977}
\begin{equation}\label{BKT}
T_\mathrm{BKT}=\frac{\pi}{2}\rho_0(T_\mathrm{BKT}).
\end{equation}
For the FF state with diagonal but anisotropic superfluid density, $\tilde\rho_{ij}=\tilde\rho_{ii}\delta_{ij}$, we have a similar $\Omega_\mathrm{w}$ but with $\tilde\omega_\mathrm{w}(\mathbf{q})=\sqrt{\sum_i\tilde\rho_{ii}q_i^2/\kappa}$. However the relation in Eq.~(\ref{BKT}) is not directly applicable for anisotropic $\tilde\rho_{ij}$. Since the BKT criterion is based on a thermodynamic argument of energy and entropy~\cite{KT}, and in the diagonal but anisotropic case the energy associated with the vortices is proportional to the geometric mean of the diagonal elements in the superfluid density tensor, i.e. $\sqrt{\Pi_i\tilde{\rho}_{ii}}$ in 2D~\cite{Williams:1998}, it is natural to expect correspondingly $T_\mathrm{BKT}=\frac{\pi}{2}\sqrt{\Pi_i\tilde{\rho}_{ii}(T_\mathrm{BKT})}$. The interplay between the FF state and the BKT phase transition is one of the main interests of this paper.

\subsection{Thermodynamic Potential and Equations}\label{Sec-Omega}

The total thermodynamic potential $\Omega=\Omega_\mathrm{s}+\Omega_\mathrm{w}=(S_\mathrm{s}+S_\mathrm{w})/\mathcal{V}$ is given by
\begin{align}
\Omega=&-\frac{1}{\mathcal{V}}\sum_\mathbf{k}\{\ln[2\cosh(\beta E_\mathbf{QK})+2\cosh(\beta h_\mathbf{QK})]-\beta\xi_\mathbf{QK}\}\nonumber\\ &+\frac{\Delta_0^2}{g}+\frac{1}{\mathcal{V}}\sum_\mathbf{q}\ln\left[1-e^{-\beta\tilde\omega_\mathrm{w}(\mathbf{q})}\right].
\end{align}
In this expression the assumption of smooth and slowly varying phase fluctuation does not take into account the presence of vortices and antivortices. In general the phase fluctuations can be separated into the sum of a static vortex part and a spin-wave part \cite{Botelho:2006}, but the vortices can be assumed to be relatively few in number when $T$ is not high. Although the vortex part might be relatively more important at very low temperatures, where the spin-wave part is suppressed but a vortex lattice can be formed, the vortex contribution to the number equations can still be (typically) small. Therefore, we choose to focus on the spin-wave fluctuations only. However, we emphasize that the effect of vortices is indeed included in the present study since the BKT transition temperature given by Eq.~(\ref{BKT}) is based on the proliferation of free vortices. At $T>T_\mathrm{BKT}$, the vortex contribution will become large, which indicates the collapse of the spin-wave description.

From $\Omega$, we can obtain several equations (Eqs.~(\ref{gapeq})-(\ref{numbereqs})) to solve. The gap equation $(\partial\Omega_\mathrm{s}/\partial\Delta_0)_{\mu,\beta,h,Q}=0$, without fluctuation contribution according to the saddle-point condition,
\begin{equation}\label{gapeq}
\frac{2}{g}-\frac{1}{V}\sum_\mathbf{k}\frac{X_\mathbf{k}}{E_\mathbf{QK}}=0.
\end{equation}
When $\Delta_0=0$, there is no need to consider $\mathbf{Q}$ which is in the phase of $\Delta$. With $\Delta_0\neq0$, a non-zero $\mathbf{Q}$ means the FF state. However, as shown in Sec.~\ref{Sec-fluctuation}, the term linear in $\theta$ in the perturbative expansion of $S_\mathrm{fl}$ vanishes intrinsically. Therefore, the equation for $Q$ does not come directly from the saddle-point condition. In order to determine $Q$, there are two possible approaches.

First, by taking the FF vector as the phase part of the order parameter, which could be treated the same as the amplitude part $\Delta_0$, we can still determine $Q$ directly from the saddle-point action in the same way as $\Delta_0$ is determined from the gap equation, i.e. $(\partial\Omega_\mathrm{s}/\partial Q)_{\beta,\mu,h,\Delta_0}=0$, or explicitly
\begin{equation}\label{Qeqs}
\frac{1}{V}\sum_\mathbf{k}\left[\frac{Q}{m}-\frac{\sinh(\beta E_\mathbf{QK})\frac{\xi_\mathbf{Qk}Q}{E_\mathbf{QK}m}-\sinh(\beta h_\mathbf{QK})\frac{\mathbf{k\cdot Q}}{mQ}}{\cosh(\beta E_\mathbf{QK})+\cosh(\beta h_\mathbf{QK})}\right]=0.
\end{equation}
Note that, although it does not conflict with the fact that the term linear in $\theta$ vanishes in the expansion of the fluctuation action, Eq.~(\ref{Qeqs}) might turn out to be trivial, if its left-hand side vanishes intrinsically as a special property of the FF state. Alternatively, as we cannot obtain the constraint of $Q$ from the saddle-point condition, it is reasonable to use the minimum of $\Omega$ rather than $\Omega_\mathrm{s}$ as the criterion for $Q$. Therefore we have
\begin{equation}\label{Qeq}
(\partial\Omega/\partial Q)_{\beta,\mu,h,\Delta_0}=0.
\end{equation}
Usually the first approach is much simpler and will be used throughout this paper, but Eq.~(\ref{Qeq}) will be discussed when necessary.

In addition to these, we also have the number equations
\begin{equation}\label{numbereqs}
n=-(\partial\Omega/\partial\mu)_{\beta,h,\Delta_0,Q},\ \ \delta n=-(\partial\Omega/\partial h)_{\beta,\mu,\Delta_0,Q},
\end{equation}
where $n=n_\uparrow+n_\downarrow$ and $\delta n=n_\uparrow-n_\downarrow$ are the total particle density and the density difference, respectively. Note that the fluctuations affect the number equations. The partial derivatives are results of the standard thermodynamic relations $n=-(\partial\Omega/\partial\mu)_{\beta,h}$ and $\delta n=-(\partial\Omega/\partial h)_{\beta,\mu}$ expanded by using the chain rule and noting that the partial derivatives of $\Omega$ with respect to $\Delta_0$ or $Q$ vanish according to the saddle-point conditions. (In the way we have phrased the problem, $\partial\Omega_\mathrm{w}/\partial\Delta_0$ is not included in accordance with the saddle-point condition for the order parameter. For the partial derivative with respect to $Q$, if the constraint in Eq.~(\ref{Qeqs}) is used, then the same argument for $\partial\Omega_\mathrm{w}/\partial\Delta_0$ also applies to $\partial\Omega_\mathrm{w}/\partial Q$.) On the other hand, Diener {\it et al}. found that including more partial derivatives by forcing the gap equation to include the fluctuation term (referred to as the ``self-consistent feedback of Gaussian fluctuation on the saddle point") will either violate the Goldstone's theorem in the Cartesian representation (with fluctuation as $\Delta=\Delta_0+\eta$) or result in ultraviolet divergence in the polar representation (with $\Delta=\Delta_0e^{i\theta}$) \cite{Diener:2008}.

\section{Phase diagram of 2D Fermi gases}\label{Sec-phasediagram}

For our aim to examine the BKT phase transition of an imbalanced system with the FF ansatz, we have to specify some details more concretely. As the BKT phase transition appears in 2D systems, the 2D contact-interaction coupling constant is renormalized like (see, e.g. Ref.~\cite{Botelho:2006})
\[\frac{1}{g}=\frac{1}{V}\sum_\mathbf{k}\frac{1}{2\epsilon_\mathbf{k}+E_b},\]
where $E_b$ is the 2D binding energy (taken as positive) of a two-particle bound state, which can be related to the 2D $s$-wave scattering length $a_s=\hbar/\sqrt{mE_b}$. The two spacial dimensions will be denoted as $x$ and $z$, then
\[\Omega_\mathrm{w}=\frac{1}{\mathcal{V}}\sum_\mathbf{q}\ln\left(1-e^{-\beta\sqrt{\frac{\tilde\rho_{xx}}{\kappa}q_x^2+\frac{\tilde\rho_{zz}}{\kappa}q_z^2}}\right)\]
with the explicit expressions
\begin{align}
\tilde\rho_{xx}=&\frac{1}{V}\sum_\mathbf{k}\left[\frac{1}{4m}\left(1-\frac{\xi_\mathbf{Qk}}{E_\mathbf{Qk}}X_\mathbf{k}\right)-\frac{\beta Y_\mathbf{k}k_x^2}{4m^2}\right],\nonumber\\
\tilde\rho_{zz}=&\frac{1}{V}\sum_\mathbf{k}\left[\frac{1}{4m}\left(1-\frac{\xi_\mathbf{Qk}}{E_\mathbf{Qk}}X_\mathbf{k}\right)-\frac{\beta Y_\mathbf{k}k_z^2}{4m^2}-3Z_\mathbf{k}k_zQ\right]\nonumber\\&-\frac{9\kappa Q^2}{4m^2}.\nonumber
\end{align}

It turns out that in the continuum limit the 2D integral in $\Omega_\mathrm{w}$ can be carried out explicitly, with a result
\[\Omega_\mathrm{w}=\frac{-\zeta(3)\kappa}{2\pi\beta^3\sqrt{\tilde\rho_{xx}\tilde\rho_{zz}}},\]
where $\zeta$ is the Riemann zeta function. It is clear that a meaningful spin-wave-like phase fluctuation requires that both $\tilde\rho_{xx}$ and $\tilde\rho_{zz}$ are positive ($\kappa$ is positive definite according to Eq.~(\ref{kappa})). This is quite natural since with negative superfluid density in either direction, the fluctuation in the corresponding mode can proliferate to decrease the energy of the system, such that any negative superfluid density results in the dynamical instability.

We can solve Eqs.~(\ref{gapeq})-(\ref{numbereqs}) self-consistently with given $T$, $E_b$, and $\delta n$ as input parameters. However, it is easier to calculate with fixed $h$, as we then do not need to solve the equation for $\delta n$. In the end it is simple, if required, to map the $h$-dependent results to the $\delta n$-dependent ones. For the numerical calculations we choose the particle mass as $m=1/2$ and the total particle density $n=1/2\pi$ such that the 2D Fermi energy $E_F=2\pi n/2m=1$.

\subsection{Without the FF State}

As is known the FF(LO) state, if it exists, often occupies only a very narrow region of the parameter space. Therefore, we start the calculation with $Q=0$. In this case the angle dependence in momentum integrations can be removed, which reduces the numerical complexity. Then also $\tilde\rho_{ij}=\tilde\rho_0\delta_{ij}$ is isotropic, with
\[\tilde\rho_0=\int\frac{kdk}{2\pi}\left[\frac{1}{4m}\left(1-\frac{\xi_\mathbf{Qk}}{E_\mathbf{Qk}}X_\mathbf{k}\right)-\frac{\beta Y_\mathbf{k}k^2}{8m^2}\right]_{Q=0},\]
and $\Omega_\mathrm{w}$ reduces to $\Omega_\mathrm{w}=-\zeta(3)\kappa/(2\pi\beta^3\tilde\rho_0)$.

Before proceeding to numerical calculations, we first clarify the phase structure qualitatively. The phase diagram is determined by the minimum of the thermodynamic potential. At high $T$ and small $E_b$, pairing is not favored, and the minimum of $\Omega$ lies at $\Delta=0$, which we refer as $\Omega_\mathrm{N}$, and the system is in the normal phase~(NP). With decreasing $T$ or increasing $E_b$, the minimum is at non-zero $\Delta$, and the pairing sets in. At the MF level, the phase diagram can be qualitatively understood by a small-$\Delta_0$ expansion of $\Omega$ around the phase transition,
\begin{equation}\label{Omegaexp}
\Omega=\Omega_\mathrm{N}+a\Delta_0^2+b\Delta_0^4+\mathcal{O}(\Delta_0^6),
\end{equation}
where $a$ and $b$ are functions of the system parameters obtained as $a=\frac{1}{2}\frac{\partial^2\Omega}{\partial\Delta_0^2}\big|_{\Delta_0=0}$ and $b=\frac{1}{24}\frac{\partial^4\Omega}{\partial\Delta_0^4}\big|_{\Delta_0=0}$. 

\subsubsection{Mean-Field Results}

First we consider the easier MF case by neglecting the fluctuations. In the balanced case,
\[b=\int d^2k\left\{\frac{\mathrm{sech}^2(\beta\xi_\mathbf{Qk}/2)[\sinh(\beta\xi_\mathbf{Qk})-\beta\xi_\mathbf{Qk}]}{16\xi_\mathbf{Qk}^3}\right\}_{Q=0}\]
is positive definite. On the other hand, $a$ changes from positive to negative continuously with decreasing $T$ or increasing $E_b$. When $a>0$, the minimum is at $\Delta_0=0$, i.e. the normal state; while for $a<0$, the minimum starts to deviate from the normal state so that $\Delta_0\approx\sqrt{-a/2b}$. Such a phase transition into paired states takes place at $a=0$ and is continuous.

\begin{figure}
\includegraphics[trim = 0 13 0 0, clip, width=0.82\columnwidth]{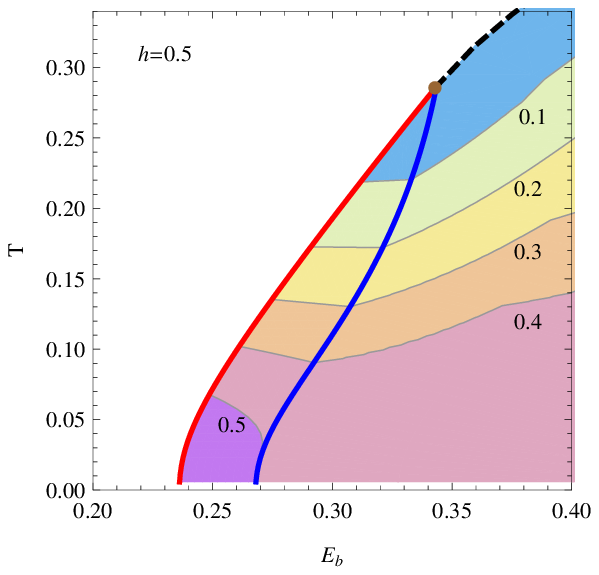}
\includegraphics[width=0.8\columnwidth]{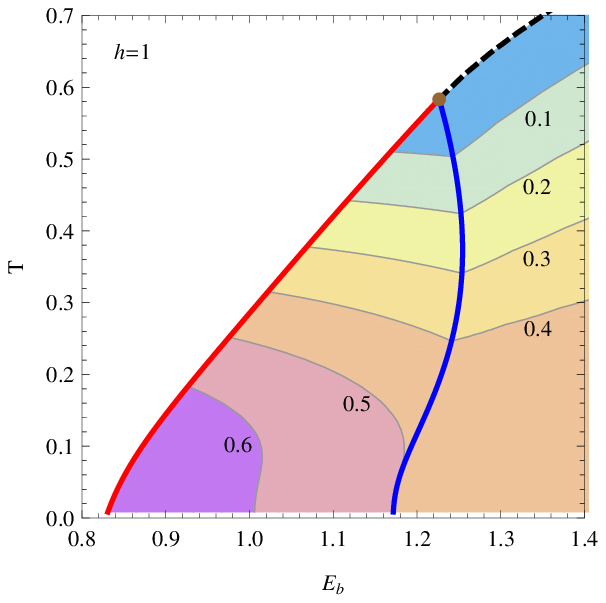}
\caption{(Color online) Mean field phase diagrams as functions of $E_b$ and $T$ without the FF ansatz at $h=0.5$ and $h=1$, respectively. The phase boundaries plotted as solid curves correspond to first-order phase transitions, while the dashed curves correspond to continuous phase transitions. The tricritical point is indicated by a brown dot where three different phases meet, i.e. the normal phase~(NP)~(white), the phase separation~(PS) region (between the red and the blue curves), and the superfluid~(SF) phase. The contours show the values of superfluid density (in the units of total density, $n=1/2\pi$) in the PS region and the SF phase, which is positive definite and approaches $n/2$ as $T\rightarrow0$ in the SF phase, since then all particles are fully paired. In the PS region it can be larger than $n/2$ because the superfluid only takes part of the spatial volume. We emphasize that the superfluid density shown here is a MF result and its non-zero value does not necessarily mean superfluidity. The phase boundaries agree with the results in Ref.~\cite{Tempere:2009}.}\label{phasediagramMF}
\end{figure}

The imbalanced case is more complicated as $b$ can become negative at large $h$. In this case higher order coefficients are positive and guarantee that the minimum of $\Omega$ is at finite $\Delta_0$. With negative $b$, if $a\leq0$, the gap equation has only one non-trivial solution corresponding to the global minimum of $\Omega_\mathrm{s}$, and all the particles are paired with non-zero $\Delta_0$ as the BCS state. In order to conform to usual terminology we call it simply the superfluid~(SF) state or phase, although strictly speaking superfluidity implies non-zero superfluid density and phase coherence rather than just non-vanishing gap parameter. But if $a>0$, the gap equation may have two non-trivial solutions, with the smaller one corresponding to a local maximum and the larger one to a local minimum. If $b$ is sufficiently negative, this local minimum can be lower than $\Omega_\mathrm{N}$ and becomes the global minimum. This phase transition taking place at non-zero $\Delta_0$ is of first-order. Such a possibility begins at the point where both $a$ and $b$ vanish, i.e. the tricritical point~\cite{Parish:2007}. The MF phase diagrams are shown in Fig.~\ref{phasediagramMF}.

\begin{figure}
\includegraphics[width=0.8\columnwidth]{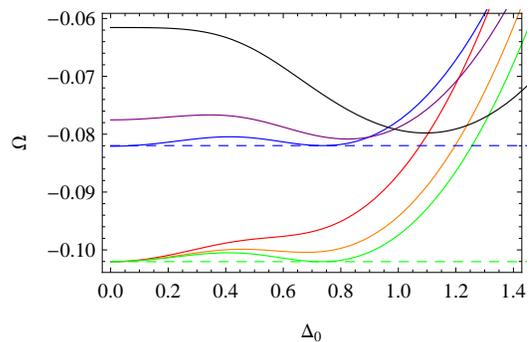}
\caption{(Color online) The thermodynamic potential $\Omega$ (with $\Omega_\mathrm{w}$ included) as functions of $\Delta_0$ at $h=0.5$ and $T=0.1$, with $E_b=0.2$ (red) for the NP with only one minimum at $\Delta_0=0$; $E_b=0.24$ (orange) for the NP with an unstable local minimum at $\Delta_0\approx0.68$; $E_b\approx0.26$ (green) for the NP-PS boundary where two minima $\Omega_\mathrm{N}$ and $\Omega(\Delta_0\approx0.73)$ are equal; $E_b\approx0.29$ (blue) for the PS-SF boundary with two equal minima $\Omega_\mathrm{N}$ and $\Omega(\Delta_0\approx0.74)$; $E_b=0.35$ (purple) for the SF phase with $\Delta_0\approx0.82$ while $\Omega_\mathrm{N}$ becomes a local minimum, and $E_b=0.6$ (black) for the SF phase with $\Delta_0\approx1.1$ as the only one minimum. Note that when all the particles are in the NP, $\mu$ is completely determined by $T$ and $h$ (here $\mu\approx1.0$), consequently $\Omega_\mathrm{N}$ is constant.}\label{Omega}
\end{figure}

When $T$ is below the tricritical value, there is a region of phase-separation~(PS) where no solution satisfying both the gap and the number equations can be found. In fact, the NP and the SF phase coexist there. The ratio of particles in these two phases is constrained by the total number density. This PS region has one boundary with the pure NP where all the particles stay unpaired, and another boundary with the pure SF phase where all the particles are paired. Between these two boundaries, the two minima of $\Omega$ remain the same, as required by the phase equilibrium condition of both phases having the same pressure. In Fig.~\ref{Omega} the curves of $\Omega(\Delta_0)$ demonstrate these cases explicitly. Here we plot the total thermodynamic potential instead of the saddle-point value. It turns out that at the temperature $T=0.1$ the effect of fluctuations is so small that the contribution to $\Omega$ is very small (cf. the boundaries shown in Fig.~\ref{phasediagramFL}). In this sense Fig.~\ref{Omega} is a useful reference for both the present and the next subsections since the way to determine the boundaries of phase-separation region is the same with and without fluctuations. We emphasize that, although our qualitative discussion about the phase transition used small-$\Delta_0$ expansion, all the numerical results presented here and hereafter are based on full calculations of the thermodynamic potential for each case.

\subsubsection{Including Fluctuations}

The above arguments are qualitatively valid when contributions from the phase fluctuations are included. Obviously, the phase fluctuations in the order parameter should not change $\Omega_\mathrm{N}$ as $\rho$ and $\Omega_\mathrm{w}$ vanish in the NP. However for the NP-PS boundary, the inclusion of the phase fluctuations for the paired states may cause a history-dependent behavior: the boundary depends on from which phase the system approaches it. Because the existence of pairs is the premise of the phase fluctuation (in our model that focuses on phase, not amplitude fluctuations), if the system starts from the NP side, all particles are unpaired such that no contribution from fluctuation should be included, and the boundary condition is $\Omega_\mathrm{N}=\Omega_\mathrm{s}(\Delta_0)$, which is exactly the MF case. However, if the boundary is approached from the PS region, the fluctuation contribution to the SF state is present since the pairs already exist. Then the equilibrium requires $\Omega_\mathrm{N}=\Omega(\Delta_0)$. Whether or not we include $\Omega_\mathrm{w}$ gives rise to different NP-PS boundary. Because $\Omega_\mathrm{w}$ is negative definite, $\Omega(\Delta_0)<\Omega_\mathrm{s}(\Delta_0)$, the NP-PS boundary obtained by $\Omega_\mathrm{N}=\Omega(\Delta_0)$ lies at smaller $E_b$ or higher $T$ compared to the MF case, and the difference increases at larger $T$ as $\Omega_\mathrm{w}$ becomes more significant. As all the pairs break up across the boundary, the disappearance of fluctuation contribution results in a sudden increase from $\Omega(\Delta_0)$ to $\Omega_\mathrm{s}(\Delta_0)$. 

However, if the fluctuation contribution to the thermodynamic potential happens to be positive (distinct from the spin-wave-like fluctuation which is negative definite), there will not be such history dependence. In that case the fluctuation makes the paired state less favored and the boundary is always obtained by $\Omega_\mathrm{N}=\Omega(\Delta_0)$, which lies at larger $E_b$ or lower $T$. On the other hand, the PS-SF boundary is independent of how it is approached, unless there is some contributions to $\Omega_\mathrm{N}$ which changes discontinuously across this boundary.

Theoretically such a sudden change of $\Omega$ across the boundary would be quite general even if we were to consider interaction effects more carefully. Because the order parameter changes discontinuously in the first-order phase transition, the change of fluctuation contributions in one phase is also discontinuous across the boundary. It would be unlikely that this discontinuity could be exactly compensated by contributions of the other phase which is continuous across the boundary (e.g. the normal state continuing from NP to PS). Strictly speaking, in the NP amplitude fluctuations might result in pairs which would be associated with the phase fluctuations. The possibility to create pairs due to the fluctuations increases dramatically as the boundary is approached because the difference between the two local minima of $\Omega$ decreases to zero. This effect is even stronger at high temperatures. Therefore, we expect the NP-PS boundary should be determined by $\Omega_\mathrm{N}=\Omega(\Delta_0)$. However, because here we only focus on the phase fluctuations for the study of the BKT mechanism, the amplitude fluctuations are beyond the scope of this paper. Furthermore, a better treatment of the normal states including the effects of interactions will certainly modify the NP-PS and the NP-SF boundaries as well. In strongly interacting systems, proper description of the normal state can be non-trivial and various Fermi-liquid, pseudogap, etc. approaches have been developed. Such a more elaborate description of the normal phase might remove the history-dependent behavior discussed above.

\begin{figure}
\includegraphics[trim = 0 15 0 0, clip, width=1\columnwidth]{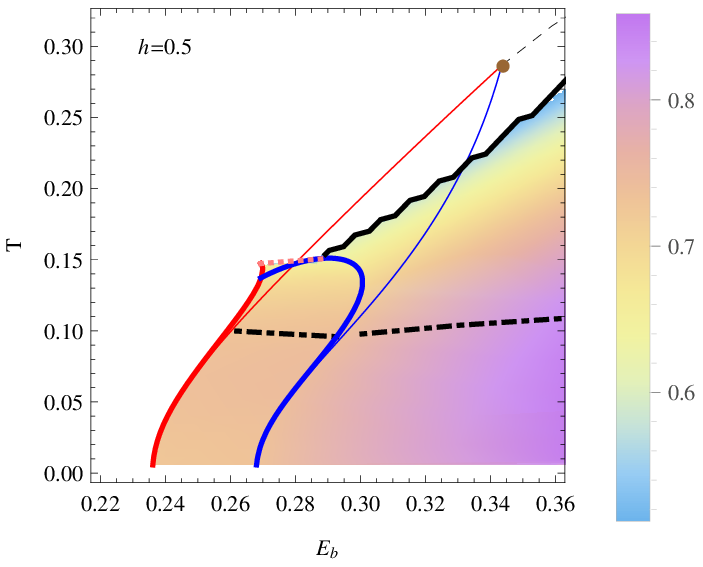}
\includegraphics[width=1\columnwidth]{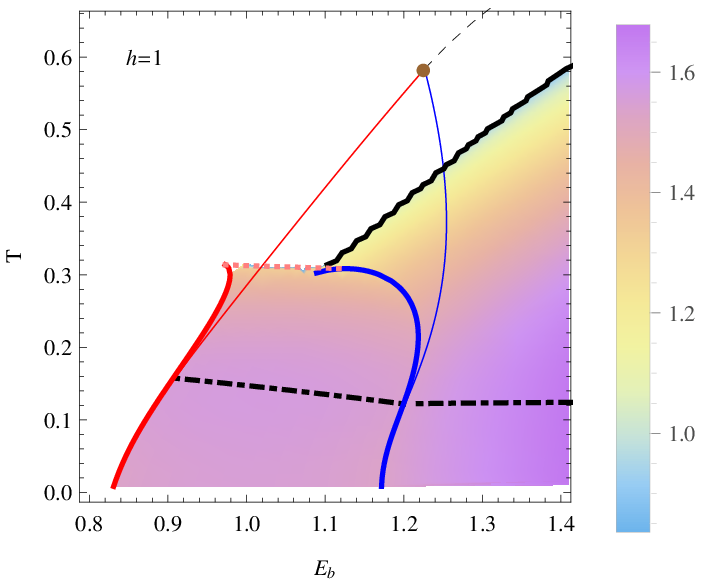}
\caption{(Color online) Phase diagrams including the fluctuations. The MF phase boundaries (thin curves) and the tricritical points are shown for comparison. The new PS-SF boundaries (thick blue) show the strong effect of fluctuations. The difference between thick and thin red curves shows the history-dependence of the NP-PS boundary. The PS region does not end with a tricritical point but with a region where no solution can be found to satisfy the equilibrium condition, as indicated by pink dotted lines. Besides the solid curves corresponding to first-order phase transitions and the dashed curves to continuous phase transitions, the black dot-dashed curves correspond to the topological BKT phase transition with $T_\mathrm{BKT}$ obtained by Eq.~(\ref{BKT}). The curves of $T_\mathrm{BKT}$ bend in the PS region as the corresponding superfluid density increases in the superfluid portion. The colored region shows the values of non-zero order parameter $\Delta_0$, but only the part below $T_\mathrm{BKT}$ can be taken as a superfluid, while the remaining part is the pseudogap phase where no phase coherence exists and the superfluid density vanishes in accordance with the BKT mechanism. The first-order NP-SF phase boundaries are not quite smooth because of the numerical difficulty to find the exact locations of this phase transition.}\label{phasediagramFL}
\end{figure}

The phase diagrams including the fluctuations are shown in Fig.~\ref{phasediagramFL}. As is clear the effect of fluctuations is significant compared to the MF results. At high $T$, a considerable region where paired states could exist in the MF case turns into pure NP due to the fact that the number equations could not be simultaneously satisfied. This region expands with increasing temperature as the fluctuations become large. Consequently, the SF phase sets in with non-zero $\Delta_0$, as can be seen from the color scales in Fig.~\ref{phasediagramFL}, thus the NP-SF phase transition becomes of first order, but note that there is no phase-coexistence at this first-order phase transition. Most interestingly, the tricritical point does not exist any longer. Instead, the PS ends with a region where we could not find any solution satisfying the equilibrium condition. Furthermore, we found the NP-PS and the PS-SF boundaries can overlap if $h$ is small. This means that, with the same $T$ and $E_b$, there can be two sets of solutions to Eq.~(\ref{gapeq}) and the phase equilibrium condition. One solution corresponds to the number constraint Eq.~(\ref{numbereqs}) satisfied in the NP, while the other to the number equation satisfied in the SF phase. 

We attribute the disappearance of the tricritical point to different fluctuation contributions to the coexisting phases. This result is distinct from the 3D case~\cite{Parish:2007}, where the tricritical point would play an important role in the phase diagram even at non-zero temperatures. In addition to the dimensionality, the main difference is that the fluctuations used in Ref.~\cite{Parish:2007} were of the Nozi{\`e}res-Schmitt-Rink~(NSR) form, which considers the pair fluctuations on the second-order phase boundary where $\Delta_0$ is small. However on the first-order boundaries, where the order parameter changes discontinuously, the NSR fluctuation is not suitable. In general, the NSR form is applicable when fluctuations are small. In this respect, the 2D and the 3D systems are different. The NSR fluctuation is widely used in 3D cases where the fluctuation is relatively weak, but the phase fluctuations which affect the first-order phase transition become much more important for the 2D cases.

\subsection{With the FF State}

Now we consider the FF state by turning on $Q$ as a free parameter. The previous case without including the FF state will be referred to as the non-FF case for the sake of simplicity. We can discuss the problem qualitatively as before by adding to Eq.~(\ref{Omegaexp}) the spatial variation of $\Delta$ as
\begin{equation}\label{OmegaexpFF}
\Omega=\Omega_\mathrm{N}+a|\Delta|^2+b|\Delta|^4+c|\nabla\Delta|^2+d|\nabla\Delta|^4+\cdots,
\end{equation}
where the expansion is up to quartic order, though even higher order expansion is possible~\cite{Combescot:2002}. With the FF ansatz $\Delta_0e^{2i\mathbf{Q}\cdot\mathbf{x}}$, the new terms correspond to an expansion in $Q$. The quadratic term $c|\nabla\Delta|^2$ plays the role of the kinetic energy of the pairs. Similar to the non-FF case, the signs of $c$ and $d$ determine the minimum of $\Omega$ along the $Q$ axis. However, as now $\Omega$ depends on both $\Delta_0$ and $Q$, a simple discussion with only one parameter is not enough. Furthermore, we find numerically that the coefficient $c$ of the total thermodynamic potential is always positive in the low temperature range of interest, which means that, unlike a 3D mass-imbalanced system~\cite{Gubbels:2009}, in the present system there is no Lifshitz point. Consequently, it is impossible to have the FF state starting from $Q=0$ and a complete calculation with $Q$ as a free parameter is necessary. We will start with the simpler case at zero temperature and then continue to the finite temperature case.

\subsubsection{Zero Temperature Limit}

Zero temperature limit, although impossible to be realized experimentally, provides clear physical insight and useful limiting behavior at low temperatures, since many calculations can be carried out analytically. At $T=0$, $\Omega_\mathrm{w}$ vanishes and $\Omega$ reduces to
\begin{align}
\Omega^{T0}=&\frac{\Delta_0^2}{g}-\frac{1}{V}\sum_\mathbf{k}[\mathrm{Max}(E_\mathbf{QK},|h_\mathbf{QK}|)-\xi_\mathbf{QK}]\nonumber\\
=&\int\frac{kdk}{2\pi}\left(\frac{\Delta_0^2}{2\epsilon_\mathbf{k}+E_b}-E_\mathbf{QK}+\xi_\mathbf{QK}\right)\nonumber\\ &+\int\frac{kdk}{2\pi}\left(\int_0^{\theta_1}+\int_{\theta_2}^\pi\right)\frac{d\theta}{\pi}(E_\mathbf{QK}-|h_\mathbf{QK}|),
\end{align}
where $\theta_{1,2}=\Re\left[\arccos\left(\frac{m(h\pm E_\mathbf{QK})}{kQ}\right)\right]$ such that $|h_\mathbf{QK}|>E_\mathbf{QK}$ is satisfied within the ranges $[0,\theta_1)$ and $(\theta_2,\pi]$. Here $\Re$ means taking the real part. It is easy to find that, as $Q\rightarrow0$, $\theta_1\rightarrow0$ and $\theta_2\rightarrow\pi\Theta(E_k-h)$ with $E_k=\sqrt{\xi_k^2+\Delta_0^2}$, $\xi_k=\frac{k^2}{2m}-\mu$ and $\Theta$ being the Heaviside step function. The first integral, being angle-independent and analytically integrable, integrates to $\frac{m}{8\pi}\left[2\Delta_0^2\ln\frac{\xi_Q+E_Q}{E_b}-\left(\xi_Q-E_Q\right)^2\right]$. As $Q\rightarrow0$, the non-FF expression for $\Omega^{T0}$ is consistent with the result in Ref.~\cite{Tempere:2007}.

\begin{figure}
\includegraphics[trim = 29 27 1 2, clip, width=0.49\columnwidth]{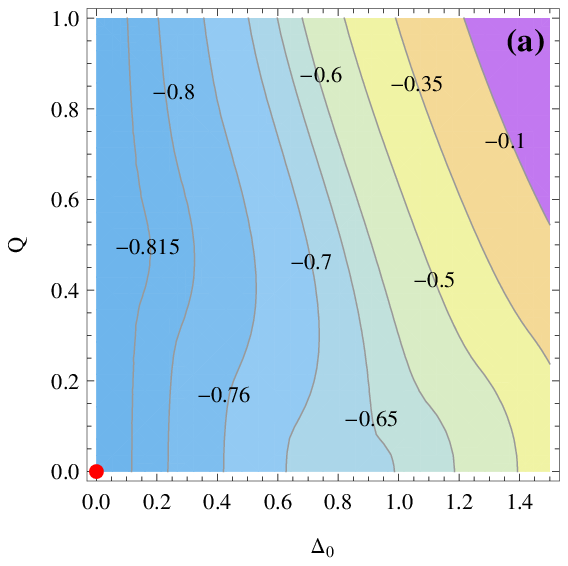}
\includegraphics[trim = 28.7 27 1.3 2, clip, width=0.49\columnwidth]{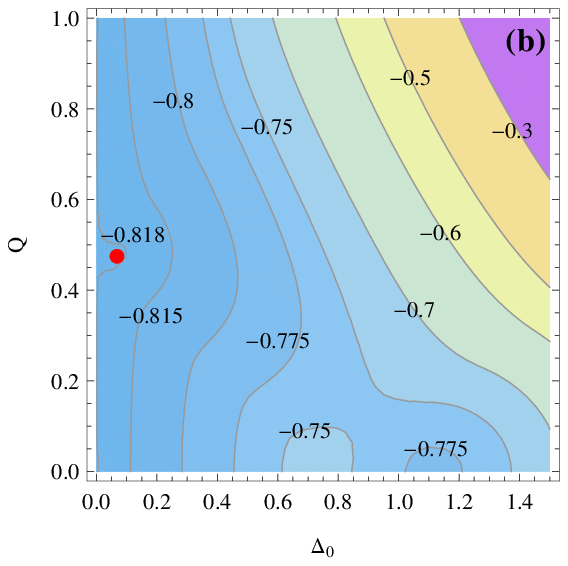}
\includegraphics[trim = 29 27 1 2, clip, width=0.49\columnwidth]{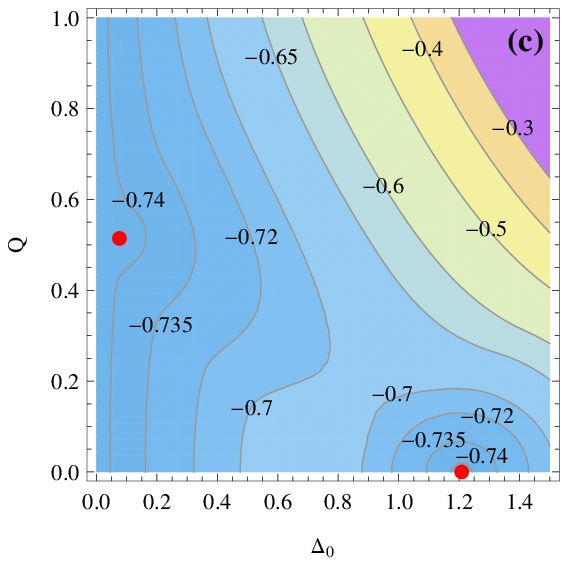}
\includegraphics[trim = 28.7 27 1.3 2, clip, width=0.49\columnwidth]{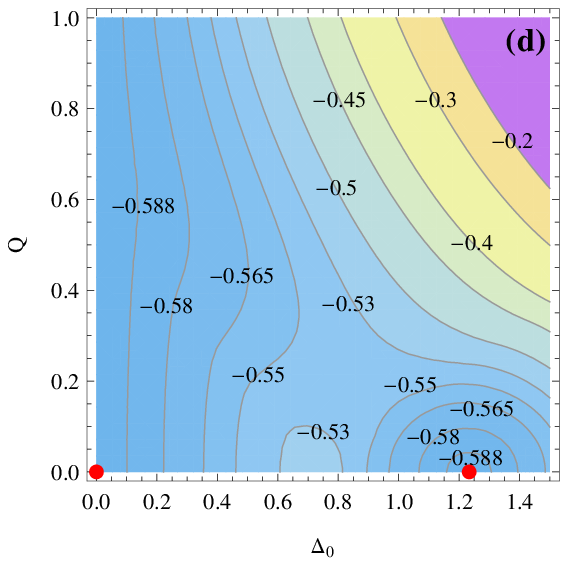}
\includegraphics[width=0.59\columnwidth]{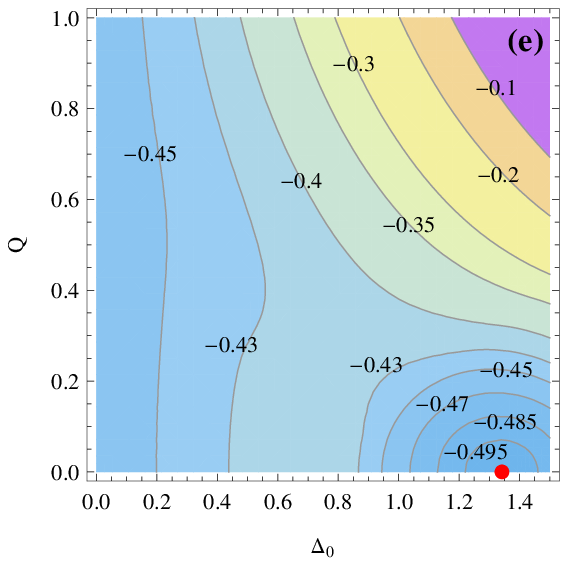}
\caption{(Color online) Contour plots of $\Omega^{T0}$ at $h=0.8$ for various $E_b$ corresponding to different phases at $T=0$. (a) $E_b=0.3$ (NP); (b) $E_b=0.5$ (FF); (c) $E_b=0.6$ (PS$_\mathrm{F}$); (d) $E_b=0.7$ (PS$_\mathrm{N}$); (e) $E_b=0.9$ (SF). For the acronyms of the phases see the text. All the axes have the same scale as in (e). The global minima are indicated by red dots. Similar results have been shown in Ref.~\cite{Conduit:2008}. The pit close to the $Q$-axis in (a) indicates the emergence of a FF state in (b) at finite $Q$. The contour labels are the values of $\Omega^{T0}$ divided by $n=1/2\pi$, i.e. the thermodynamic potential per particle.}\label{OmegaContour}
\end{figure}

\begin{figure}
\includegraphics[width=1\columnwidth]{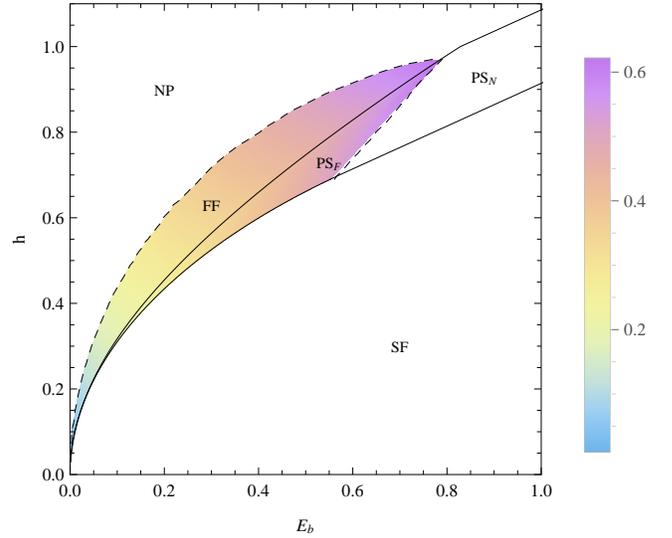}
\caption{(Color online) Phase diagram of a 2D imbalanced Fermi gas at $T=0$. Here NP is the normal phase, PS$_\mathrm{F}$ is the phase separation region with the FF and the superfluid~(SF) states coexisting, while in PS$_\mathrm{N}$ the SF phase and the NP coexist. The phase boundaries plotted as solid curves correspond to first-order phase transitions, while the dashed curves correspond to continuous phase transitions. The colors in the FF-related regions show the values of $Q$ of the FF states.}\label{phasediagramT0}
\end{figure}

The phase diagram is determined by the global minimum of $\Omega^{T0}$ in the $\Delta_0$-$Q$ plane. There can be three different local minima, namely $\Omega_\mathrm{N}$ of the normal state with $\Delta_0=0$, $\Omega_\mathrm{SF}$ of the paired state with $\Delta_0\neq0$ but $Q=0$, and $\Omega_\mathrm{FF}$ of the FF state with both $\Delta_0$ and $Q$ non-zero, each of which can become the global minimum depending on the parameters. Coexistence is possible between the SF and the FF phases, as well as between the SF phase and the NP. Such coexistence is not possible between the NP and the FF phase since in the cases we have studied $\Omega_\mathrm{FF}$ is always lower than $\Omega_\mathrm{N}$ when the FF state exists. This issue has been discussed more extensively in Ref.~\cite{Conduit:2008}. Fig.~\ref{OmegaContour} shows various examples of the contour plots of the thermodynamic potential $\Omega^{T0}$. It should be noted that the FF state sets in with infinitesimal $\Delta_0$ but non-vanishing $Q$. However, any state with $\Delta_0=0$ should be taken as the normal state since a non-zero $Q$ has no contribution when $\Delta_0=0$. Therefore, the corresponding NP-FF phase transition is still continuous, which is different from the non-FF case and not associated with a negative coefficient $c$ in Eq.~(\ref{OmegaexpFF}). The complete phase diagram at $T=0$ is shown in Fig.~\ref{phasediagramT0}, from which we see the FF state exists in a horn-shaped area and gives way to the normal state when $h$ or $E_b$ becomes large, resulting in two parts of the PS region: one as the coexistence of the FF and the SF phases~(PS$_\mathrm{F}$) at smaller $h$ and $E_b$, and the other of the NP and the SF phase~(PS$_\mathrm{N}$). This phase diagram can be taken as the generalization of the previous results of 2D imbalanced Fermi gases in homogeneous case \cite{He:2008} or in lattices \cite{Kujawa-Cichy:2011}. While these studies did not consider FFLO states, they found similar phase boundaries as we do in Fig.~\ref{phasediagramT0} with their partially polarized phases replaced by our FF phases at small imbalance.

The superfluid density at $T=0$ can also be calculated. It is important to note that
\begin{align}
\tilde{\rho}_{xx}^{T0}&=\frac{n}{4m}-\int\frac{kdk}{2\pi}\frac{k(\sin\theta_2+\sin\theta_1)}{4mQ\pi}\equiv\frac{\partial_Q\Omega^{T0}}{4Q},\nonumber
\end{align}
which means that the FF state whose $Q$ satisfies $\partial_Q\Omega^{T0}=0$ always has a superfluid density tensor with a vanishing component along the direction perpendicular to the FF vector. This property of the transverse superfluid density (stiffness) has been pointed out in Refs.\cite{Radzihovsky:2009,Radzihovsky:2011} based on the GL theory and a symmetry argument. It means that there is no energy cost to generate fluctuations along the $x$ direction, which can be understood from the divergence of $\Omega_\mathrm{w}$ as $\tilde{\rho}_{xx}=0$ in the denominator. It is not a serious problem at $T=0$ as the thermal fluctuation is not considered, however a vanishing or small superfluid density will cause difficulties when we use the spin-wave description of the phase fluctuations at finite temperatures.

\subsubsection{Finite Temperature}

\begin{figure*}
\includegraphics[width=0.67\columnwidth]{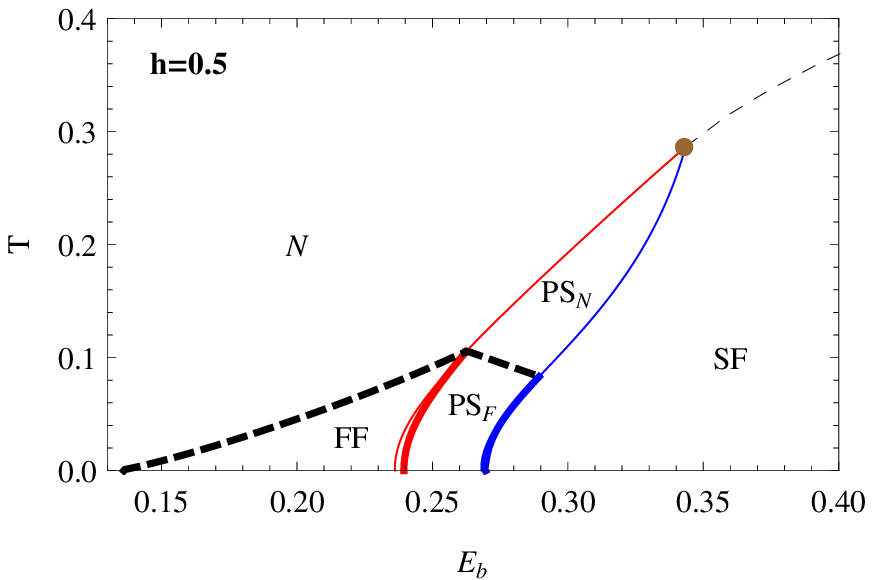}
\includegraphics[width=0.67\columnwidth]{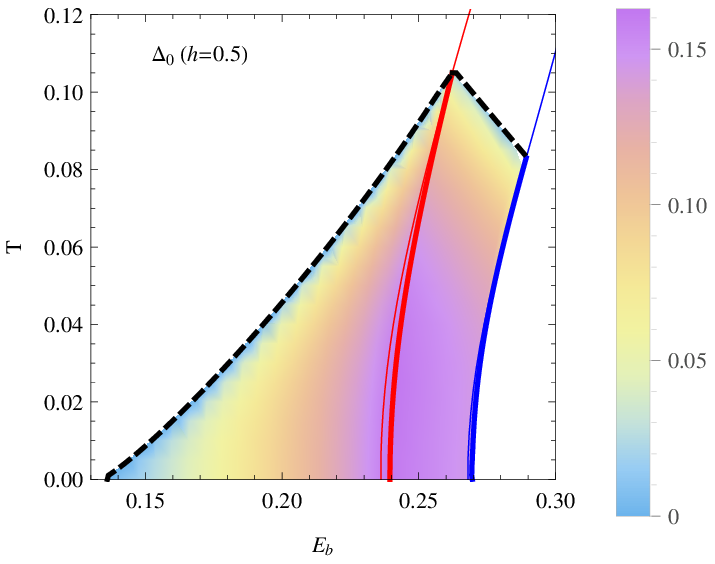}
\includegraphics[width=0.67\columnwidth]{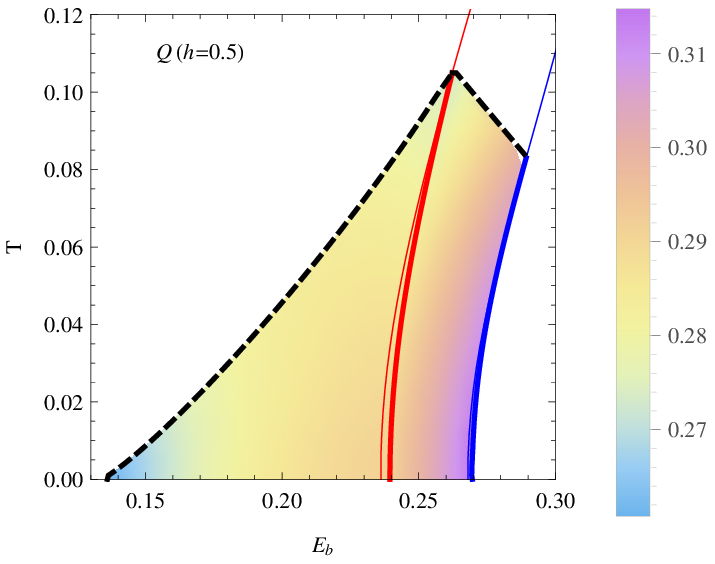}
\includegraphics[width=0.67\columnwidth]{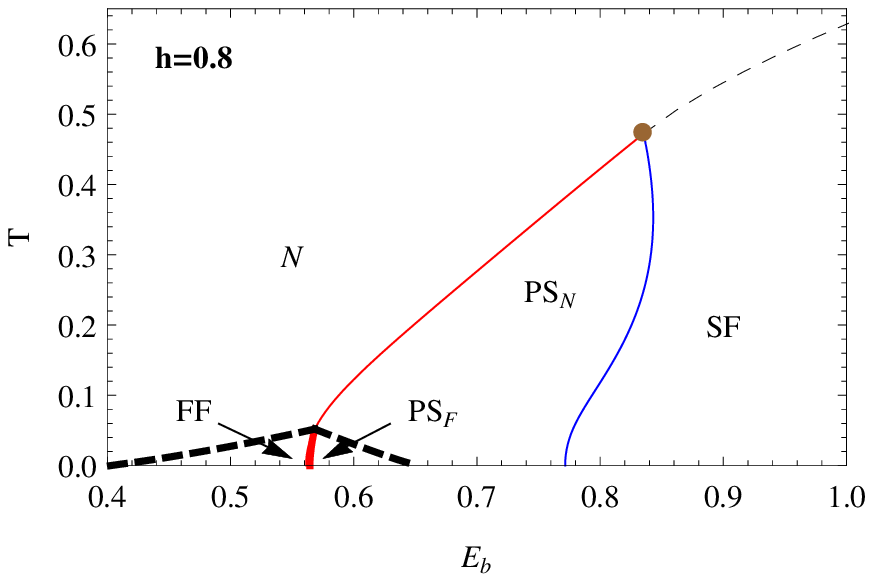}
\includegraphics[width=0.67\columnwidth]{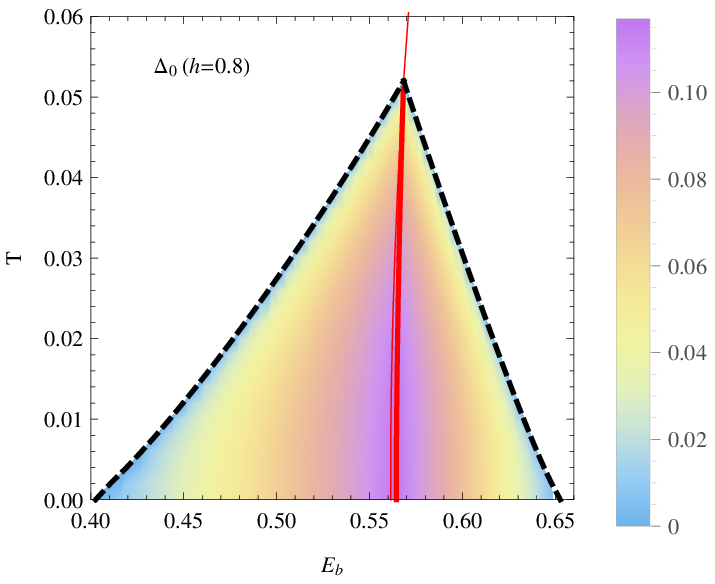}
\includegraphics[width=0.67\columnwidth]{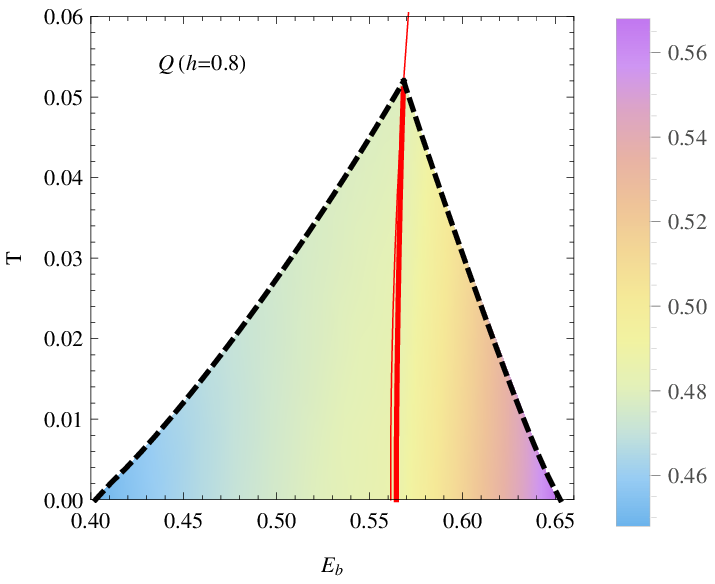}
\caption{(Color online) Left: Mean field phase diagrams as functions of $E_b$ and $T$ with the FF ansatz at $h=0.5$ and $h=0.8$, respectively. The new boundaries after including the FF ansatz are shown by thick curves, while the phase boundaries (thin curves) and the tricritical points in the corresponding non-FF case are shown for comparison. Since in the case with $h=1$ the FF state does not exist even at zero temperature (cf. Fig.~\ref{phasediagramT0}), the case with $h=0.8$ is used instead, where the PS$_\mathrm{N}$ region exists even at $T=0$. The phase boundaries plotted as solid curves correspond to first-order phase transitions, while the dashed curves correspond to continuous phase transitions. Middle and Right: The density plots for the values of $\Delta_0$ and $Q$ of the FF states, zoomed in for the FF-related regions.}\label{MFFF}
\end{figure*}

Similar to the non-FF case, we first present the MF results with the FF ansatz in Fig.~\ref{MFFF}, which can be taken as finite-temperature extensions of the results of Fig.~\ref{phasediagramT0}. Compared to the non-FF cases, the PS$_\mathrm{F}$ regions are shifted a bit towards the SF-phase side and also shrink slightly. On the MF level, the range of $E_b$ with the possibility of FF state shrinks smoothly with increasing $T$, and at higher temperatures the FF states can survive around the FF-PS$_\mathrm{F}$ boundaries, where the peaks of $\Delta_0$ are located (but in general the values of $Q$ increase with $E_b$). However, in order to draw more reliable conclusions we must include the fluctuations for such a 2D system.

\begin{figure}
\includegraphics[width=0.8\columnwidth]{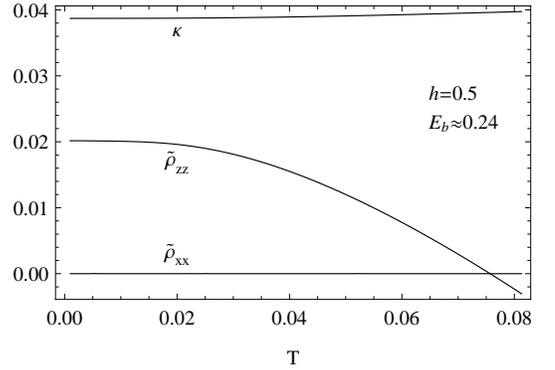}
\caption{$\kappa$, $\tilde{\rho}_{xx}$ and $\tilde{\rho}_{zz}$ of the FF state as functions of $T$. The parameters are chosen from the FF-PS$_\mathrm{F}$ boundary at $T=0$. $T$ ranges from $0$ to where the FF state can still be found.}\label{kappa-rhoFF}
\end{figure}

Being aware of the vanishing transverse superfluid density at $T=0$, we first check the behavior of the superfluid density $\tilde{\rho}$ at non-zero temperatures, which has a significant effect on the phase fluctuations. Fig.~\ref{kappa-rhoFF} shows the $T$-dependence of $\tilde{\rho}_{xx}$ and $\tilde{\rho}_{zz}$, as well as $\kappa$ given in Eq.~(\ref{kappa}), of a FF state, where we see that $\tilde{\rho}_{xx}$ always vanishes and $\tilde{\rho}_{zz}$ can become negative at high $T$, while $\kappa$ is positive definite. In fact, we find numerically that the relation $\tilde{\rho}_{xx}=\frac{\partial_Q\Omega_\mathrm{s}}{4Q}$ is still true at finite temperature, such that the FF state always has divergent transverse fluctuations if the FF vector is determined by Eq.~(\ref{Qeqs}), i.e. the condition $(\partial\Omega_\mathrm{s}/\partial Q)_{\beta,\mu,h,\Delta_0}=0$. It is interesting to explore this property from another angle. Starting with the vanishing transverse superfluid density of the FF ansatz, which may be argued based on symmetry, the relation $\partial_Q\Omega_\mathrm{s}=4Q\tilde{\rho}_{xx}$ means that the left-hand side of Eq.~(\ref{Qeqs}) vanishes identically for a FF state. Then the absence of terms linear in $\theta$ in the expansion of $S_\mathrm{fl}$ is a natural consequence: Since the transverse fluctuations are unconstrained, it is physically justified to be unable to determine the value of $Q$ from the saddle-point condition. Therefore, all these special properties of the FF ansatz are connected.

\begin{figure}
\includegraphics[width=0.8\columnwidth]{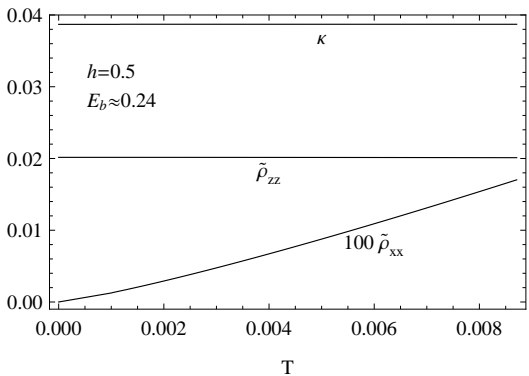}
\caption{The same as Fig.~\ref{kappa-rhoFF} but with $Q$ determined by Eq.~(\ref{Qeq}). The transverse superfluid density $\tilde{\rho}_{xx}$ is magnified $100$ times for the sake of clarity.}\label{kappa-rhoFF-PSFfullQ}
\end{figure}

In this respect, we try to determine $Q$ by Eq.~(\ref{Qeq}) rather than (\ref{Qeqs}), and the results are shown in Fig.~\ref{kappa-rhoFF-PSFfullQ}. While $\tilde{\rho}_{xx}$ is then non-zero it is still quite small. As a rough estimate, if we take $\tilde{\rho}_{xx}$ depending on $T$ linearly at low temperatures, where the non-zero $\tilde{\rho}_{zz}$ and $\kappa$ change very little, then $\Omega_\mathrm{w}\propto\frac{T^3\kappa}{\sqrt{\tilde{\rho}_{xx}\tilde{\rho}_{zz}}}$ is approximately proportional to $T^{5/2}$, which vanishes at $T=0$ but increases very fast with $T$. Such an increase is significant also due to the small coefficient of the proportionality $\tilde{\rho}_{xx}(T)\sim1.5\times10^{-2}T$. For this reason we do not expect this different approach to change our conclusions considerably.

\begin{figure*}
\includegraphics[width=0.73\columnwidth]{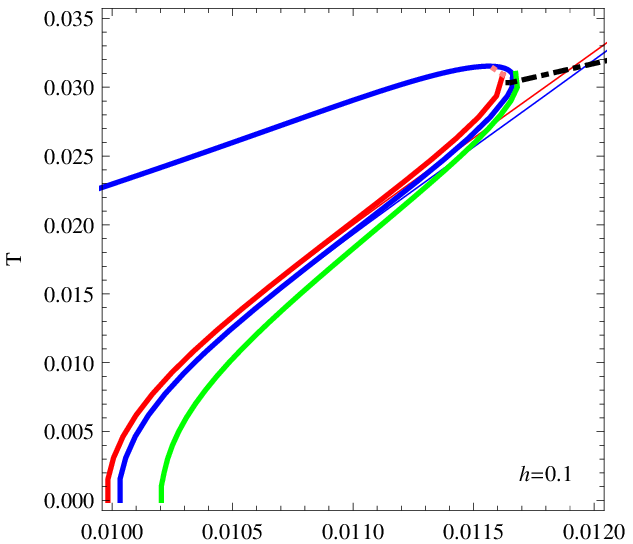}
\includegraphics[width=0.66\columnwidth]{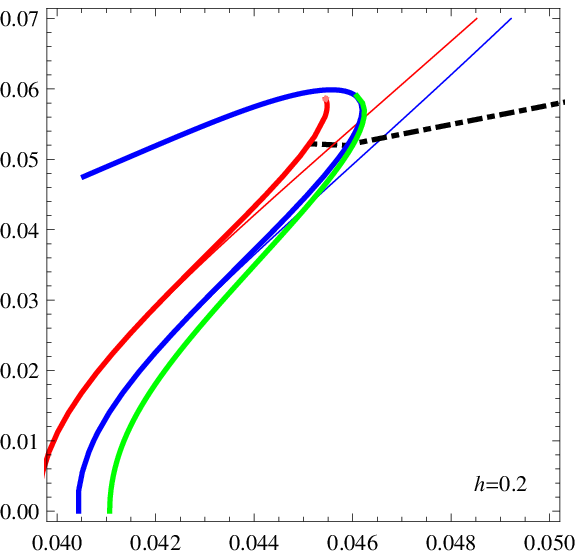}
\includegraphics[width=0.66\columnwidth]{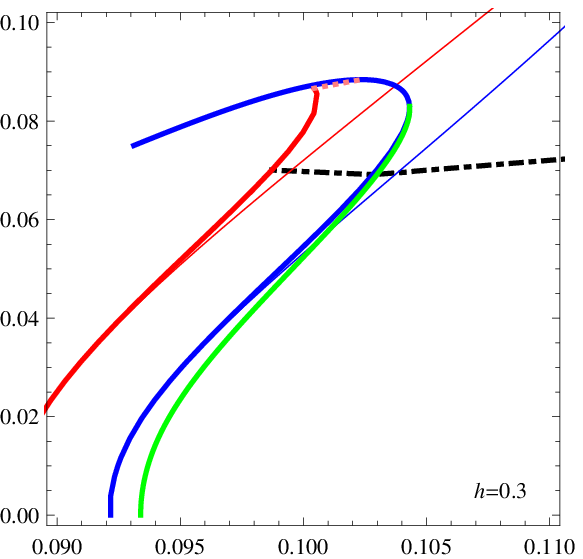}
\includegraphics[width=0.72\columnwidth]{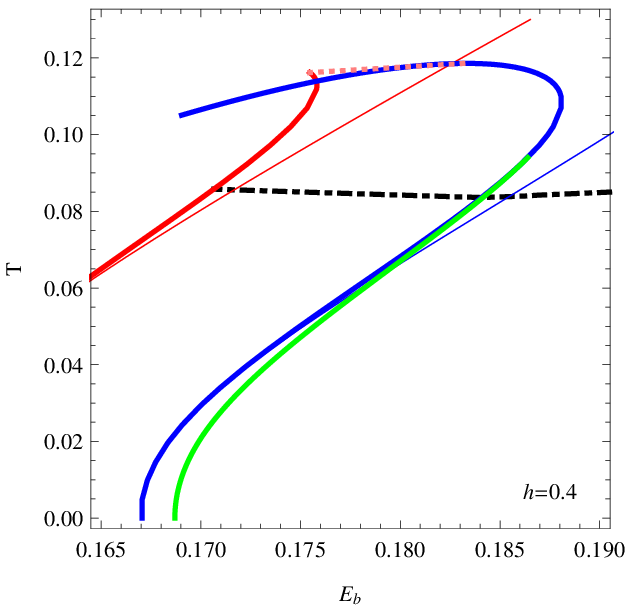}
\includegraphics[width=0.67\columnwidth]{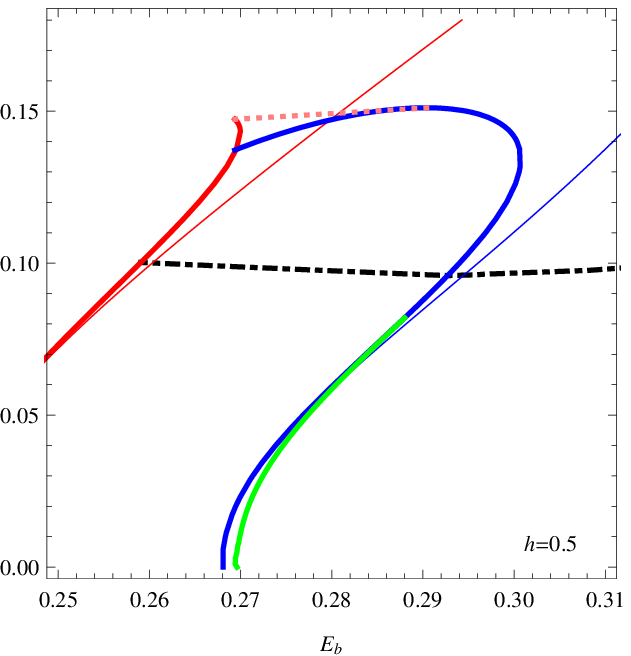}
\includegraphics[width=0.66\columnwidth]{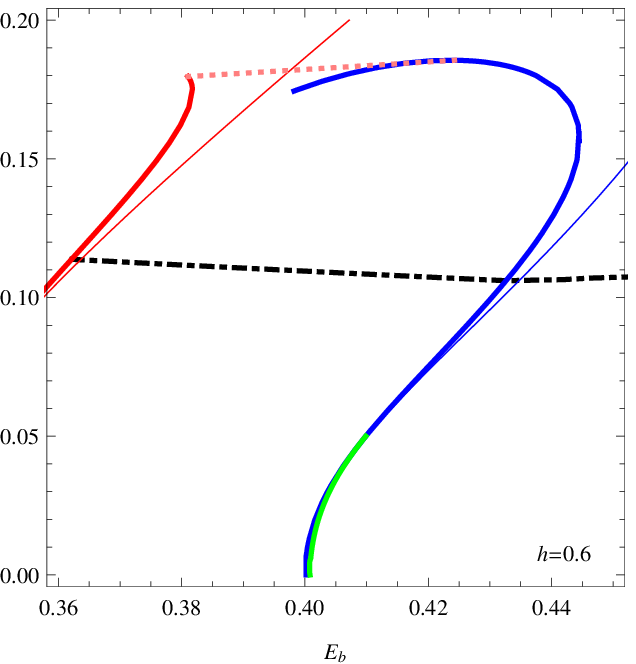}
\caption{(Color online) By including the FF state, newly obtained PS$_\mathrm{F}$-SF segments (thick green) for various $h$ are added to the phase diagrams of the corresponding non-FF cases. The non-FF results are plotted as in Fig.~\ref{phasediagramFL}, namely with the red curves for the NP-PS boundaries and the blue curves for the PS-SF boundaries, in which the thin curves are the MF results and the thick ones include the fluctuations. Also, the pink dotted lines indicate the regions where no solution satisfies the phase equilibrium condition, and the black dot-dashed curves are $T_\mathrm{BKT}^\mathrm{SF}$ obtained in the non-FF case. Because the PS$_\mathrm{F}$-SF segments are quite close to the PS-SF boundaries of the non-FF case, we only focus on the relevant part in each plot to demonstrate the difference clearly. In these plots $h$ ranges from $0.1$ to $0.6$ as the PS$_\mathrm{F}$-SF boundary does not exist when $h\geq0.7$.}\label{phasediagramQ}
\end{figure*}

Because of the divergent fluctuations, it is impossible to determine part of the finite temperature phase diagram where the fluctuation contributions to the FF state should be included, such as the NP-FF and the FF-PS$_\mathrm{F}$ boundaries. However, the PS$_\mathrm{F}$-SF boundary does not have such a numerical difficulty since the FF state is empty and its phase fluctuations do not need to be included. Despite the incompleteness, we still present our results in Fig.~\ref{phasediagramQ} for various $h$. Different from the non-FF case, the PS-SF boundary now starts with a PS$_\mathrm{F}$-SF segment at low temperatures, which lies to the right of the corresponding PS-SF boundary in the non-FF case. Then this PS$_\mathrm{F}$-SF segment gradually approaches the latter, and finally merges into it as the FF state gives way to the normal state. We find that for $h=0.2,0.3$, and $0.4$, the PS$_\mathrm{F}$-SF boundaries extend above the corresponding $T_\mathrm{BKT}$ obtained in the isotropic non-FF case. This suggests that the effect of anisotropic superfluidity might be relevant to the BKT mechanism. 

It should be pointed out that in the PS$_\mathrm{F}$ region there are two superfluid densities associated with the FF~($\tilde{\rho}^\mathrm{FF}$) and the SF phases~($\tilde{\rho}^\mathrm{SF}$). Correspondingly there are two critical temperatures $T_\mathrm{BKT}^\mathrm{FF}$ and $T_\mathrm{BKT}^\mathrm{SF}$, respectively. Here $\tilde{\rho}^\mathrm{SF}$ is isotropic and qualitatively the same as the non-FF case, while for $\tilde{\rho}^\mathrm{FF}$ the criterion should be $T_\mathrm{BKT}^\mathrm{FF}=\frac{\pi}{2}\sqrt{\tilde{\rho}_{xx}^\mathrm{FF}(T_\mathrm{BKT}^\mathrm{FF})\tilde{\rho}_{zz}^\mathrm{FF}(T_\mathrm{BKT}^\mathrm{FF})}$. Since $\tilde{\rho}_{xx}^\mathrm{FF}=0$, $T_\mathrm{BKT}^\mathrm{FF}$ would be zero (or almost zero, if there can be some mechanisms to suppress the marginally divergent fluctuation of the FF state, e.g. finite-size effects or broken symmetries). Even if $Q$ is determined from the full thermodynamic potential, see Fig.~\ref{kappa-rhoFF-PSFfullQ}, we can estimate $T_\mathrm{BKT}^\mathrm{FF}$ to be less than $10^{-3}$.

Because of the very strong phase fluctuations from the vanishing $\tilde{\rho}_{xx}^\mathrm{FF}$, we expect the FF state to be destroyed at very low $T$. Since above $T_\mathrm{BKT}$ any quasi-long-range phase coherence could not survive, a constant $Q$ in the phase of a plane-wave ansatz characterizing the FF state is not consistent with the BKT mechanism. Consequently it is likely that the FF-PS$_\mathrm{F}$ boundary will probably be replaced by the NP-PS boundary of the non-FF case. This is supported by the fact that the difference between $\Omega_\mathrm{FF}$ and $\Omega_\mathrm{N}$ is quite small, and the order parameter $\Delta_0$ of the FF state~($\Delta_\mathrm{FF}$) is not very large. Therefore, it is reasonable to expect that the FF state will be easily replaced by the normal state when the fluctuations are strong. Meanwhile, the isotropic SF phase will behave similarly as in the non-FF case, and $T_\mathrm{BKT}^\mathrm{SF}$ should behave as $T_\mathrm{BKT}$ in Fig.~\ref{phasediagramFL}. Then the behavior of the system will be the same as in the non-FF case.

Strictly speaking, $T_\mathrm{BKT}$ sets a threshold for the fluctuation where the assumption of the smooth fluctuations of a spin-wave form, or equivalently, the small-$q$ expansion of $S_\mathrm{fl}$, turns out invalid due to the proliferation of free vortices which destroy the (quasi-)~long-range order and the phase coherence. As was predicted by Nelson and Kosterlitz~\cite{Nelson:1977} in the isotropic case the superfluid density jumps from $\rho=2T_\mathrm{BKT}/\pi$ to zero as $T$ crosses $T_\mathrm{BKT}$ from below. This has been observed experimentally in 2D $^4$He films~\cite{Bishop:1978} and recently also in cold Bose gases~\cite{Noh:2013}. We expect it is also true for the anisotropic FF state. On this account, above $T_\mathrm{BKT}$ the action in Eq.~(\ref{flaction}) is no longer of the spin-wave form. Since the FF state is unstable at finite temperature due to the fluctuations, the PS$_\mathrm{F}$-SF boundaries shown in Fig.~\ref{phasediagramQ} are also vulnerable, consequently the region between the PS$_\mathrm{F}$-SF boundary (thick green curves) and the PS-SF boundary (thick blue curves) in the non-FF case would be affected by the instability. On the PS$_\mathrm{F}$ side across the PS$_\mathrm{F}$-SF boundary where particles start to occupy the FF state, the fluctuations at finite temperatures will destroy the FF state, and a new equilibrium between the NP and the SF phase is established instead.

As a final remark, taking the BKT mechanism into account, our results with the fluctuations above $T_\mathrm{BKT}$ shown in the phase diagrams are not quantitatively reliable because the small-$q$ expansion becomes less reliable, although qualitatively they still give some useful information. Our calculations already show that the regions of the phase diagrams with paired states are reduced significantly from the MF results at high $T$ due to fluctuations. In order to draw more quantitative conclusions, a more complete calculation of the thermal fluctuations at higher temperatures is required, e.g. with the original fluctuation action in Eqs.~(\ref{Gauss-fl-action}) and (\ref{Dij}). In addition, throughout our calculation the NP is taken as a free Fermi gas. This could be improved by describing it as a Fermi liquid~\cite{Baym:1991} which would lower the energy of the normal phase. We expect that this difference can modify the phase diagrams quantitatively, but not change them qualitatively.

\section{Summary and Discussions}

By studying the phase diagram of 2D imbalanced Fermi gases based on the thermodynamic potential on the MF level, we find the existence of the FF state at zero temperature. The possibility of FF state at finite temperatures and its effect on the BKT mechanism are discussed by including phase fluctuations. We also obtained the superfluid density tensor for the anisotropic FF state which always has vanishing transverse component. 

The effect of the phase fluctuations is demonstrated, which turn out to be very strong for the FF state and possibly destroy the FF-related phases at finite temperatures. Therefore, it would be quite hard to experimentally observe the FF state in nearly infinite continuum 2D Fermi gases, unless extremely low temperatures can be achieved. Since the strong phase fluctuations destroying the quasi-long-range order and phase coherence result in a breakdown of the spin-wave approximation of the fluctuation action, an improved study of the FF state at finite temperatures should take account of the fluctuations more completely. We note, as an interesting line of research, that a dispersion relation for collective excitations including higher order terms $\propto q^4$ has been introduced for unitary Fermi gases by Salasnich {\it et al.} \cite{Salasnich:2008} and applied at the finite (low) temperature in Ref.~\cite{Salasnich:2010}. Besides, a recent experiment with Niobium nitride films~\cite{Yong:2013} has shown that the standard BKT mechanism which only considers the phase fluctuations might not be enough to accurately describe the 2D superconductor (superfluid) phase transition, and a comprehensive consideration including also the amplitude fluctuations might be necessary~\cite{Erez:2013}.

Another candidate for an inhomogeneous order parameter is the LO state which does not have the problem of a vanishing transverse superfluid density~\cite{Radzihovsky:2011}, however it was also claimed to be unstable to a nematic phase at non-zero temperatures~\cite{Radzihovsky:2009}. There has been one paper studying the BKT phase transition of the LO (stripe) state for an anisotropic 2D system composed of coupled 1D tubes~\cite{Lin:2011}, where several different BKT critical temperatures associated with different defects are discussed and found to be linearly dependent on the intertube coupling. Nevertheless, it is an open and important question whether a more general FFLO-type state can be stable against thermal fluctuations and how this might affect the BKT mechanism. In addition, other mechanisms such as optical lattices and trapping potentials can reduce the role of fluctuations because of broken symmetries~\cite{Koponen:2007PRL,Loh:2010}. Also a mass imbalance~\cite{He:2006,Conduit:2008} or spin-orbit coupling effects~\cite{Wu:2013,Liu:2013} can enhance the Fermi-surface asymmetry and increase the stability of the FFLO state. These topics will be considered in our future work.

\section*{Acknowledgements}

This work was supported by the Academy of Finland through its Centers of Excellence Program (2012-2017) and under Projects No.~263347, No.~141039, No.~251748, No.~135000, and No.~272490. This research was supported in part by the National Science Foundation under Grant No. NSF PHY11-25915.

\appendix

\section{Hubbard-Stratonovich Transformation}\label{transformation}

For the sake of clarity, let us first introduce the notations used in the appendices as well as in the main text. Within the Euclidean space-time (dimension 1+d with 1 for time and d for space), the coordinate vector is denoted as $x=(\tau,\mathbf{x})$, and the momentum as $k=(i(ik_n),\mathbf{k})$ with the Matsubara frequency $ik_n=(2n+1)\pi/\beta$ (fermionic) or $ik_n=2n\pi/\beta$ (bosonic), where $\beta=1/T$ is the inverse of temperature. However, when there is no risk of confusion between vectors and numbers, sometimes $x$ (or $k$) can also be used for the norm of $\mathbf{x}$ (or $\mathbf{k}$). The vector product in d-space is indicated as $\mathbf{k}\cdot\mathbf{x}$ while the product of space-time vectors is written as $kx$, e.g. $ikx=i\mathbf{k}\cdot\mathbf{x}-ik_n\tau$. The discrete momentum space and the continuous coordinate space are linked via the Fourier transformation and the Fourier series formulae $f(k)=\frac{1}{\mathcal{V}}\int f(x)e^{-ikx}dx$ and $f(x)=\sum_kf(k)e^{ikx}$, where $\sum_k$ includes the summation over the Matsubara frequencies as well as the momenta, and $\mathcal{V}=V\beta$ with $V$ as the total volume of the d-dimensional space. In the continuum limit the summation over the spacial momenta can be carried out as an integration.

According to the standard Hubbard-Stratonovich transformation, a bosonic field operator $\hat\Delta(x)$ is introduced via the functional integral relation $1\propto\int\mathcal{D}\hat\Delta^*\mathcal{D}\hat\Delta e^{-\int dx[\hat\Delta^*(x)-g\hat\psi^\dagger_\uparrow(x)\hat\psi^\dagger_\downarrow(x)](1/g)[\hat\Delta(x)-g\hat\psi_\downarrow(x)\hat\psi_\uparrow(x)]}$ which is inserted to the microscopic partition function $Z=\int\mathcal{D}\hat\psi^\dagger_\uparrow\mathcal{D}\hat\psi_\uparrow\mathcal{D}\hat\psi^\dagger_\downarrow\mathcal{D}\hat\psi_\downarrow e^{-S}$. Here
\begin{widetext}
\[S=\int dx\left[\sum_\sigma\hat\psi^\dagger_\sigma(x)\partial_\tau\hat\psi_\sigma(x)+\hat{H}(x)\right]=\int dxdx'\left[-\sum_\sigma\hat\psi^\dagger_\sigma(x)G^{-1}_{0\sigma}(x,x')\hat\psi_\sigma(x')-g\hat\psi^\dagger_\uparrow(x)\hat\psi^\dagger_\downarrow(x')\hat\psi_\downarrow(x')\hat\psi_\uparrow(x)\delta(x-x')\right],\]
where $G^{-1}_{0\sigma}(x,x')=(-\partial_\tau-\hat\varepsilon+\mu_\sigma)\delta(x-x')$ is the inverse of a free fermion propagator for species $\sigma$. Then a new action $\tilde{S}$ in the resulting partition function $Z=\int\mathcal{D}\hat\psi^\dagger_\uparrow\mathcal{D}\hat\psi_\uparrow\mathcal{D}\hat\psi^\dagger_\downarrow\mathcal{D}\hat\psi_\downarrow\mathcal{D}\hat\Delta^*\mathcal{D}\hat\Delta e^{-\tilde{S}}$ can be written in a quadratic form,
\[\tilde{S}=\int dxdx'\left\{\frac{|\hat\Delta(x)|^2}{g}\delta(x-x')-\sum_\sigma\hat\psi^\dagger_\sigma(x)G^{-1}_{0\sigma}(x,x')\hat\psi_\sigma(x')-\left[\hat\psi^\dagger_\uparrow(x)\hat\Delta(x)\hat\psi^\dagger_\downarrow(x')+\hat\psi_\downarrow(x')\hat\Delta^*(x)\hat\psi_\uparrow(x)\right]\delta(x-x')\right\}.\]
\end{widetext}
By using the Nambu-Gorkov basis $\hat\Psi^\dagger=(\hat\psi^\dagger_\uparrow, \hat\psi_\downarrow)$ and $\hat\Psi=(\hat\psi_\uparrow, \hat\psi^\dagger_\downarrow)^T$, the action can be expressed as
\[\tilde{S}=\int dxdx'\left[\frac{|\hat\Delta(x)|^2}{g}\delta(x-x')-\hat\Psi^\dagger(x)\mathbf{G}^{-1}(x,x')\hat\Psi(x')\right],\]
where 
\[\mathbf{G}^{-1}(x,x')=\left(\begin{array}{cc}
-\partial_\tau-\hat\varepsilon+\mu_\uparrow & \hat\Delta(x)\\
\hat\Delta^*(x) & -\partial_\tau+\hat\varepsilon-\mu_\downarrow
\end{array}\right)\delta(x-x').\]
In the momentum space the action becomes
\[\tilde{S}=\mathcal{V}\sum_q\frac{|\hat\Delta(q)|^2}{g}-\mathcal{V}\sum_{k,k'}\hat\Psi^\dagger(-k)\mathbf{G}^{-1}(k,k')\hat\Psi(k'),\]
with
\begin{align}
\mathbf{G}^{-1}&(k,k')=\nonumber\\
&\left(\begin{array}{cc}
(ik'_n-\epsilon_\mathbf{k'}+\mu_\uparrow)\delta_{k,k'} & \hat\Delta(k-k')\\
\hat\Delta^*(-k+k') & (ik'_n+\epsilon_\mathbf{k'}-\mu_\downarrow)\delta_{k,k'}
\end{array}\right)\nonumber
\end{align}
as the Fourier transform of $\mathbf{G}^{-1}(x,x')$.

Integrating out the Fermi fields, we get the effective bosonic action
\begin{align}\label{eff}
S_\mathrm{eff}&=\int dx\frac{|\hat\Delta(x)|^2}{g}-\mathrm{Tr}\ln[-\beta\mathbf{G}^{-1}(x,x')]\nonumber\\
&=\mathcal{V}\sum_{iq_n,\mathbf{q}}\frac{|\hat\Delta(q)|^2}{g}-\mathrm{Tr}\ln[-\beta\mathbf{G}^{-1}(k,k')],
\end{align}
where $\mathrm{Tr}$ means the trace over the Nambu space as well as the (1+d) coordinate or momentum space. Since for a matrix operation $\mathrm{tr}\ln=\ln\mathrm{det}$ (here $\mathrm{tr}$ means only the trace in the Nambu space) and $\mathbf{G}^{-1}$ is a $2\times2$ matrix, the minus sign inside the logarithm makes no difference and will be dropped for simplicity. 

Now the original functional integral of Fermi fields has been transformed into an integral over the Bose field $\hat\Delta$. However, since the action is a complicated function of $\hat\Delta$, in general it cannot be carried out explicitly unless some approximation is made. A widely used one is the MF approximation, also referred to as the saddle-point method, which is a good approximation if the fields vary smoothly and no strong correlation is present. In MF approximation the integral over the field $\hat\Delta$ is replaced by using its expectation value $\langle\hat\Delta\rangle=\Delta_\mathrm{s}$. This parameter is also referred to as the order parameter, and it satisfies the saddle-point condition $\delta S_\mathrm{s}/\delta\Delta^*_\mathrm{s}=0$. For a constant $\Delta_\mathrm{s}$, $\Delta_\mathrm{s}(k-k')=\Delta_\mathrm{s}\delta_{k,k'}$ and $\mathbf{G}_\mathrm{s}^{-1}\equiv\mathbf{G}^{-1}(\Delta_\mathrm{s})$ is diagonal in momentum space. In this case the functional integral reduces to $Z\propto e^{-S_\mathrm{s}}$ with~\cite{Stoof:2009,Tempere:2007}
\begin{equation}\label{saddle}
S_\mathrm{s}\equiv S_\mathrm{eff}(\Delta_\mathrm{s})=\frac{\mathcal{V}|\Delta_\mathrm{s}|^2}{g}-\sum_k\ln[\mathrm{det}\beta\mathbf{G}_\mathrm{s}^{-1}(k)].
\end{equation}

In general $\Delta_\mathrm{s}$ might not be constant, and in momentum space $\mathbf{G}^{-1}$ might not be diagonal. This may cause some problems, especially when we need to invert $\mathbf{G}^{-1}$ into $\mathbf{G}$. However, for the FF ansatz $\Delta_\mathrm{s}(x)=\Delta_0e^{2i\mathbf{Q}\cdot\mathbf{x}}$ whose Fourier transform is $\Delta_\mathrm{s}(k)=\Delta_0\delta_{\mathbf{k},2\mathbf{Q}}$, the coordinate-dependent phase can be removed by shifting the momenta of $\hat\psi_\uparrow(\mathbf{k})$ and $\hat\psi_\downarrow(\mathbf{k})$ into $\mathbf{Q+k}$ and $\mathbf{Q-k}$, respectively, which automatically means that the total momentum is $2\mathbf{Q}$. This shift is a special case of the gauge transformation in Eq.~(\ref{gaugetransform}). The resulting $\tilde{\mathbf{G}}_\mathrm{s}^{-1}$ becomes diagonal as
\begin{align}
\tilde{\mathbf{G}}_\mathrm{s}^{-1}(k,k')&=\tilde{\mathbf{G}}_\mathrm{s}^{-1}(k)\delta_{k,k'}\nonumber\\
=&\left(\begin{array}{cc}
ik_n-\epsilon_\mathbf{Q+k}+\mu_\uparrow & \Delta_0\\
\Delta_0 & ik_n+\epsilon_\mathbf{Q-k}-\mu_\downarrow
\end{array}\right)\delta_{k,k'},\nonumber
\end{align}
and $\Delta_\mathrm{s}$ reduces to $\Delta_0$.

\section{Fluctuation Action Obtained from the Saddle-Point Action}\label{action-fl}

In Appendix~\ref{transformation} the derivation of the saddle-point action does not involve fluctuations. For an arbitrary form of the order parameter, the inverse Nambu propagator $\mathbf{G}_\mathrm{s}^{-1}$ is usually not diagonal, which hinders the derivation of the explicit expression of the action. The FF ansatz is a very special case for which the momentum shift makes $\tilde{\mathbf{G}}_\mathrm{s}^{-1}$ diagonal. In this case the derivation is almost the same as in the case of a constant $\Delta_\mathrm{s}$. One cannot expect a simple shift or transformation for an order parameter with random fluctuations. However, as will be seen below, the small-$q$ expansion used in Sec.~\ref{Sec-phasefl} to obtain the fluctuation action in a spin-wave form actually relaxes the momentum constraint, e.g. $\delta_{k-k',q}$ in Eq.~(\ref{perturbationK}), which is the source of the problematic off-diagonal terms. With the small-$q$ expansion it becomes possible to generalize the saddle-point calculation to include smooth fluctuations.

First we note that there is a way to simplify the expression of the full inverse Nambu propagator $\tilde{\mathbf{G}}^{-1}$. Since in Eq.~(\ref{perturbationK}) the $\theta$-dependence appears only in the diagonal terms of $\tilde{\mathbf{K}}$, it can be absorbed into the chemical potentials in $\tilde{\mathbf{G}}^{-1}_\mathrm{s}$~\cite{Tempere:2009}. Then we can split $\tilde{\mathbf{G}}^{-1}$ into $\bar{\mathbf{G}}^{-1}_\mathrm{s}+\bar{\mathbf{K}}$, where $\bar{\mathbf{K}}(k,k')=\eta(k-k')\sigma_1$ and $\bar{\mathbf{G}}^{-1}_\mathrm{s}$ is simply $\tilde{\mathbf{G}}^{-1}_\mathrm{s}$ with $\mu_\sigma$ replaced by $\bar\mu_\sigma$,
\begin{align}
\bar\mu_\uparrow&=\mu_\uparrow-\frac{i\partial_\tau\theta}{2}+\frac{i(\nabla\theta\cdot\nabla_\mathbf{Q}+\frac{1}{2}\nabla_\mathbf{Q}\cdot\nabla\theta)}{2m}-\frac{(\nabla\theta)^2}{8m},\nonumber\\
\bar\mu_\downarrow&=\mu_\downarrow-\frac{i\partial_\tau\theta}{2}-\frac{i(\nabla\theta\cdot\nabla_\mathbf{-Q}+\frac{1}{2}\nabla_\mathbf{-Q}\cdot\nabla\theta)}{2m}-\frac{(\nabla\theta)^2}{8m}.\nonumber
\end{align}
Their Fourier transforms are
\begin{widetext}
\begin{align}
\bar\mu_\uparrow(k,k')&=\mu_\uparrow\delta_{k,k'}+\sum_q\left[-\frac{q_n\theta(q)}{2}-\frac{i\theta(q)}{4m}(\mathbf{k}^2-\mathbf{k'}^2+3\mathbf{q}\cdot\mathbf{Q})\right]\delta_{k-k',q}+\sum_{q,q'}\frac{\theta(q)\theta(q')\mathbf{q}\cdot\mathbf{q}'}{8m}\delta_{k-k',q+q'},\nonumber\\
\bar\mu_\downarrow(k,k')&=\mu_\downarrow\delta_{k,k'}-\sum_q\left[\frac{q_n\theta(q)}{2}-\frac{i\theta(q)}{4m}(\mathbf{k}^2-\mathbf{k'}^2-3\mathbf{q}\cdot\mathbf{Q})\right]\delta_{k-k',q}+\sum_{q,q'}\frac{\theta(q)\theta(q')\mathbf{q}\cdot\mathbf{q}'}{8m}\delta_{k-k',q+q'}.
\end{align}
\end{widetext}
Note that it would seem as if by absorbing the $\theta$-dependent terms into the chemical potentials, we do not only simplify the perturbative matrix $\bar{\mathbf{K}}$ but also loosen the requirement that $\theta$ should be smooth in space-time. Furthermore, if we only consider the phase fluctuations, the perturbative part $\bar{\mathbf{K}}$ vanishes and the remaining part $\bar{\mathbf{G}}^{-1}_\mathrm{s}$ is in a saddle-point form. However, this simplification is only superficial since it moves the difficulties into $\bar{\mathbf{G}}^{-1}_\mathrm{s}$, because $\bar\mu_\sigma$ is no longer a c-number but an operator which involves off-diagonal terms in the momentum space.

In small-$q$ expansion, we assume that the fluctuations of $\theta$ change much more smoothly and slowly than the Fermi fields, so that the functional integral over the Fermi fields can be carried out adiabatically. In this way the momentum (or position) of $\theta$ is no longer associated with the Fermi fields since the field $\theta$ can be taken as a constant, and the difficulty of the off-diagonal terms no longer exists. It then becomes possible to carry out independent Fourier transformations of the Fermi fields without involving $\theta(x)$ and we get
\begin{align}
\bar\mu_\uparrow&=\mu_\uparrow-\frac{i\partial_\tau\theta}{2}+\frac{i[i\nabla\theta\cdot(\mathbf{k+Q})+\frac{1}{2}\nabla_\mathbf{Q}\cdot\nabla\theta]}{2m}-\frac{(\nabla\theta)^2}{8m},\nonumber\\
\bar\mu_\downarrow&=\mu_\downarrow-\frac{i\partial_\tau\theta}{2}-\frac{i[i\nabla\theta\cdot(\mathbf{k-Q})+\frac{1}{2}\nabla_\mathbf{-Q}\cdot\nabla\theta]}{2m}-\frac{(\nabla\theta)^2}{8m},\nonumber
\end{align}
which are diagonal in momentum space. From these we define
\begin{align}\label{shiftmu}
\bar\mu&=\mu-\frac{i\partial_\tau\theta}{2}-\frac{\nabla\theta\cdot\mathbf{Q}}{2m}+i\frac{\nabla_\mathbf{Q}\cdot\nabla\theta-\nabla_\mathbf{-Q}\cdot\nabla\theta}{8m}-\frac{(\nabla\theta)^2}{8m},\nonumber\\
\bar h&=h-\frac{\nabla\theta\cdot\mathbf{k}}{2m}+i\frac{\nabla_\mathbf{Q}\cdot\nabla\theta+\nabla_\mathbf{-Q}\cdot\nabla\theta}{8m}.
\end{align}
Since these ``barred" chemical potentials can be taken as c-numbers during the fermionic functional integral, using $\bar\mu_\sigma$ instead of $\mu_\sigma$ will not change the derivation for the saddle-point action in Appendix~\ref{transformation}. Now with the phase fluctuations only, the results in Eqs.~(\ref{saddle}) and (\ref{saddleaction}) can be directly generalized by Eq.~(\ref{shiftmu}), that is, we get $\bar S_\mathrm{s}$ with $\{\bar\mu,\bar h\}$ replacing $\{\mu,h\}$ in $S_\mathrm{s}$. To complete this adiabatic approximation of the functional integral, we have to introduce an extra integral over the coordinate of $\theta$, divided by $\mathcal{V}$ to insure correct dimensions. It means that we use the space-time average of the fluctuations corresponding to the long-wavelength and low-frequency limit. Keeping the quadratic order of the derivatives of $\theta$ in the expansion of $\bar S_\mathrm{s}$, we get $\bar S_\mathrm{s}=S_\mathrm{s}+\bar S_\mathrm{fl}$ with $S_\mathrm{s}$ given in Eq.~(\ref{saddle}) (or Eq.~(\ref{saddleaction})) and
\begin{align}\label{barSfl}
\bar S_\mathrm{fl}=&\frac{\mathcal{V}}{2}\int\frac{dx}{\mathcal{V}}\left[\kappa\left(\frac{\partial\theta}{\partial\tau}\right)^2+\rho_{ij}\nabla_i\theta\nabla_j\theta+B_{++}(\nabla_\mathbf{Q}\cdot\nabla\theta)^2\right.\nonumber\\ &\quad\left.+B_{--}(\nabla_\mathbf{-Q}\cdot\nabla\theta)^2+B_{+-}(\nabla_\mathbf{Q}\cdot\nabla\theta)(\nabla_\mathbf{-Q}\cdot\nabla\theta)\vphantom{\frac{1}{2}}\right.\nonumber\\ &\left.+(\mathbf{A}_+\cdot\nabla\theta)\nabla_\mathbf{Q}\cdot\nabla\theta+(\mathbf{A}_-\cdot\nabla\theta)\nabla_\mathbf{-Q}\cdot\nabla\theta\vphantom{\left(\frac{}{}\right)^2}\right],
\end{align}
where
\begin{align}
\kappa=&\frac{1}{V}\sum_\mathbf{k}\frac{\Delta_0^2X_\mathbf{k}+\beta E_\mathbf{Qk}\xi_\mathbf{Qk}^2Y_\mathbf{k}}{4E_\mathbf{Qk}^3},\nonumber\\
\rho_{ij}=&\frac{1}{V}\sum_\mathbf{k}\left[\frac{\delta_{ij}}{4m}\left(1-\frac{\xi_\mathbf{Qk}}{E_\mathbf{Qk}}X_\mathbf{k}\right)-\frac{\beta Y_\mathbf{k}k_ik_j}{4m^2}\right.\nonumber\\&\left.\qquad\qquad-Z_\mathbf{k}(k_iQ_j+Q_ik_j)\vphantom{\frac{1}{2}}\right]-\frac{\kappa Q_iQ_j}{m^2}.\nonumber
\end{align}
These generalize the results in Ref.~\cite{Tempere:2009} by including the FF ansatz. $\mathbf{A}_\pm$ and $B_{s_1s_2}$ are complicated functions of $\mu$, $h$, $\beta$, $\Delta_0$ and $\mathbf{Q}$, and the summation over spacial indices $i$ and $j$ is assumed, the factor $\mathcal{V}/2$ is taken out for later convenience. Besides, some terms linear in $\partial_\tau\theta$, $\nabla\theta$, $\nabla_\mathbf{\pm Q}\cdot\nabla\theta$ and their mixed products $\partial_\tau\theta\nabla\theta$, $\partial_\tau\theta\nabla_\mathbf{\pm Q}\cdot\nabla\theta$ are omitted since their contributions vanish after the overall integral (for $\partial_\tau\theta$, note that $\theta$ is bosonic so $\int_0^\beta d\tau\partial_\tau\theta=0$ due to the periodic boundary condition). 

The presence of the FF vector makes the expression of $\bar S_\mathrm{fl}$ very complicated. If $\mathbf{Q}=0$, we find that the contribution from $\mathbf{A}_\pm$ vanishes after the overall integral and the one from $B_{s_1s_2}$ corresponds to higher order correction (quartic in momentum after Fourier transformation), so only the first two terms survive. But with $\mathbf{Q}\neq0$, there are lower order contributions from $\mathbf{A}_\pm$ and $B_{s_1s_2}$ which are relevant. These can be calculated by using the following Fourier transformations
\begin{widetext}
\begin{align}
&\int\frac{dx}{\mathcal{V}}(\nabla\theta)\nabla_\mathbf{\pm Q}\cdot\nabla\theta=\sum_{q,p}\int\frac{dx}{\mathcal{V}}[\nabla\theta(q)e^{iqx}]\nabla_\mathbf{\pm Q}\cdot\nabla\theta(p)e^{ipx}=\sum_{q,p}i\mathbf{q}\theta(q)i(\mathbf{p\pm Q})\cdot i\mathbf{p}\theta(p)\delta_{q,-p}\approx\pm\sum_{q}i\mathbf{q}|\theta(q)|^2\mathbf{Q\cdot q},\nonumber\\
&\int\frac{dx}{\mathcal{V}}(\nabla_{s_1\mathbf{Q}}\cdot\nabla\theta)(\nabla_{s_2\mathbf{Q}}\cdot\nabla\theta)=\sum_{q,p}\int\frac{dx}{\mathcal{V}}[\nabla_{s_1\mathbf{Q}}\cdot\nabla\theta(q)e^{iqx}][\nabla_{s_2\mathbf{Q}}\cdot\nabla\theta(p)e^{ipx}]\nonumber\\ &\qquad\quad=\sum_{q,p}[i(\mathbf{q}+s_1\mathbf{Q})\cdot i\mathbf{q}\theta(q)][i(\mathbf{p}+s_2\mathbf{Q})\cdot i\mathbf{p}\theta(p)]\delta_{q,-p}=\sum_{q}[\mathbf{q}^4-s_1s_2(\mathbf{Q\cdot q})^2]|\theta(q)|^2\approx-s_1s_2\sum_{q}(\mathbf{Q\cdot q})^2|\theta(q)|^2,\nonumber
\end{align}
\end{widetext}
where we used $\theta(-q)=\theta^*(q)$ for real $\theta(x)$ and kept terms up to the quadratic order of $q$. Together with the Fourier transform of the first two terms in $\bar S_\mathrm{fl}$, we finally get
\begin{align}
\bar S_\mathrm{fl}=&\frac{\mathcal{V}}{2}\sum_q[\kappa q_n^2+\rho_{ij}q_iq_j+i(\mathbf{A}_+-\mathbf{A}_-)\cdot\mathbf{q}(\mathbf{Q\cdot q})\nonumber\\&\qquad+(B_{+-}-B_{++}-B_{--})(\mathbf{Q\cdot q})^2]|\theta(q)|^2\nonumber\\=&\frac{\mathcal{V}}{2}\sum_q(\kappa q_n^2+\tilde\rho_{ij}q_iq_j)|\theta(q)|^2,
\end{align}
where $\tilde\rho_{ij}\equiv\rho_{ij}+\mathbf{A}_iQ_j+\mathbf{A}_jQ_i+BQ_iQ_j$ with
\begin{align}
\mathbf{A}\equiv&\frac{i}{2}(\mathbf{A}_+-\mathbf{A}_-)=-\frac{\kappa}{2m^2}\mathbf{Q}-\sum_\mathbf{k}\frac{1}{2}Z_\mathbf{k}\mathbf{k},\nonumber\\
B\equiv&B_{+-}-B_{++}-B_{--}=-\frac{\kappa}{4m^2}.\nonumber
\end{align}
In conclusion we find,
\begin{align}
\tilde\rho_{ij}=&\frac{1}{V}\sum_\mathbf{k}\left[\frac{\delta_{ij}}{4m}\left(1-\frac{\xi_\mathbf{Qk}}{E_\mathbf{Qk}}X_\mathbf{k}\right)-\frac{\beta Y_\mathbf{k}k_ik_j}{4m^2}\right.\nonumber\\&\left.\qquad\qquad-\frac{3Z_\mathbf{k}}{2}(k_iQ_j+Q_ik_j)\vphantom{\frac{1}{2}}\right]-\frac{9\kappa Q_iQ_j}{4m^2},\nonumber
\end{align}
which is in general not diagonal if $\mathbf{Q}\neq0$. However we can choose the direction of $\mathbf{Q}$ as, e.g. the z-axis, then $h_\mathbf{Qk}=h-\frac{Qk_z}{m}$, such that $X_\mathbf{k}$, $Y_\mathbf{k}$ and $Z_\mathbf{k}$ are even in all spatial momentum components $k_i$ except for $k_z$ (note that $\xi_\mathbf{Qk}$ and $E_\mathbf{Qk}$ are always even in $\mathbf{k}$). Therefore, $\sum_\mathbf{k}Y_\mathbf{k}k_ik_j=\sum_\mathbf{k}Y_\mathbf{k}k_i^2\delta_{ij}$, $\sum_\mathbf{k}Z_\mathbf{k}k_i=\sum_\mathbf{k}Z_\mathbf{k}k_z\delta_{iz}$, and $\tilde\rho_{ij}$ reduces to
\begin{align}
\tilde\rho_{ij}=&\frac{1}{V}\sum_\mathbf{k}\left[\frac{\delta_{ij}}{4m}\left(1-\frac{\xi_\mathbf{Qk}}{E_\mathbf{Qk}}X_\mathbf{k}\right)-\frac{\beta Y_\mathbf{k}k_i^2\delta_{ij}}{4m^2}\right.\nonumber\\&\left.\qquad\qquad\qquad-3Z_\mathbf{k}k_zQ\delta_{iz}\delta_{jz}\vphantom{\frac{1}{2}}\right]-\frac{9\kappa Q^2\delta_{iz}\delta_{jz}}{4m^2}.\nonumber
\end{align}
This expression is diagonal, but with $\tilde\rho_{zz}$ different from other diagonal elements. The results of $\kappa$ and $\tilde\rho_{ij}$ obtained in this way are consistent with those obtained in Sec.~\ref{Sec-phasefl} by the direct small-$q$ expansion of $\mathbf{D}_{22}$ in Eq.~(\ref{Dij}). Similar derivation and results were presented in a recent paper for the 3D case \cite{Devreese:2013}. We emphasize that our derivation is generally applicable to other dimensions than two as well.

\end{document}